\documentclass[twocolumn]{aastex701}
\usepackage{CJK}
\usepackage{verbatim}
\usepackage{amsmath}
\usepackage{soul, xcolor}
\usepackage{booktabs}
\usepackage{rotating}
\usepackage{tabularx}
\usepackage{booktabs}

\newcommand{\RNum}[1]{\uppercase\expandafter{\romannumeral #1\relax}}

\newcommand{\ha}{\mbox{H$\alpha$}}
\newcommand{\hb}{\mbox{H$\beta$}}
\newcommand{\hg}{\mbox{H$\gamma$}}
\newcommand{\paa}{Pa$\alpha$}
\newcommand{\OIII}{[\ion{O}{3}]\,$\lambda\lambda$4959,\,5007}

\newdimen\digitwidth    
\setbox1=\hbox{0}       
\digitwidth=\wd1        
\catcode`"=\active      
\def"{\kern\digitwidth}

\shorttitle{ABCD of LRDs}
\shortauthors{Chen et al.}

\begin{document}
\begin{CJK*}{UTF8}{gbsn}

\title{ABCD: The Nuclear Structure of the Little Red Dots Revealted through Absorption, Break, Continuum, and Decrement}

\author[0009-0003-4721-177X]{Chang-Hao Chen (陈昌灏)}
\affiliation{Kavli Institute for Astronomy and Astrophysics, Peking University, Beijing 100871, China}
\affiliation{Department of Astronomy, School of Physics, Peking University, Beijing 100871, China}
\email[show]{cchen$\_$louis@stu.pku.edu.cn}

\author[0000-0002-4569-9009]{Jinyi Shangguan (上官晋沂)}
\affiliation{Kavli Institute for Astronomy and Astrophysics, Peking University, Beijing 100871, China}
\affiliation{Department of Astronomy, School of Physics, Peking University, Beijing 100871, China}
\email[show]{shangguan@pku.edu.cn}

\author[0000-0001-6947-5846]{Luis C. Ho}
\affiliation{Kavli Institute for Astronomy and Astrophysics, Peking University, Beijing 100871, China}
\affiliation{Department of Astronomy, School of Physics, Peking University, Beijing 100871, China}
\email{lho.pku@gmail.com}

\author[0000-0002-2420-5022]{Zijian Zhang (张子键)}
\affiliation{Kavli Institute for Astronomy and Astrophysics, Peking University, Beijing 100871, China}
\affiliation{Department of Astronomy, School of Physics, Peking University, Beijing 100871, China}
\email{zjz.kiaa@stu.pku.edu.cn}

\author[0000-0001-9840-4959]{Kohei Inayoshi}
\affiliation{Kavli Institute for Astronomy and Astrophysics, Peking University, Beijing 100871, China}
\affiliation{Department of Astronomy, School of Physics, Peking University, Beijing 100871, China}
\email{inayoshi0328@gmail.com}

\author[0000-0001-8496-4162]{Ruancun Li (李阮存)}
\affiliation{Max-Planck-Institut f\"{u}r Extraterrestrische Physik (MPE), Giessenbachstr., D-85748 Garching, Germany}
\email{liruancun@mpe.mpg.de}

\begin{abstract}
We present a spectroscopic analysis of 14 little red dots (LRDs) at redshifts $2.2 < z < 6.7$ using NIRSpec/MSA prism and medium-resolution grating observations, aiming to constrain the nuclear gas structure through Balmer emission-line profiles, absorption features, relative line intensities, and continuum properties. We simultaneously decompose the broad, narrow, and absorption components of \ha, \hb, and \hg, and measure both integrated line ratios and velocity-resolved Balmer decrements. The narrow-line Balmer decrements are broadly consistent with Case~B recombination modified by mild dust attenuation, while the broad-line decrements are elevated to levels consistent with photoionization models of high-density gas at $n_{\rm H} \gtrsim 10^9\ {\rm cm^{-3}}$. Velocity-resolved Balmer decrements in five sources with highest signal-to-noise ratio are centrally peaked. Assuming virialized broad-line region dynamics, our model can reproduce the Balmer decrement profiles in three sources using a radial density profile with a power-law index $\beta<2$. The Balmer absorption lines detected in six sources yield absorber covering factors exceeding $50\%$. Sources with blueshifted absorption lines tend to have elevated narrow-line Balmer decrement, suggesting a connection between dust content and the presence of outflow. Comparing the incident luminosity inferred from broad and narrow \ha\ emission with the continuum suggests that both the UV and optical continuum and the line emission are linked by photoionization. We propose that the distinctive spectral and continuum properties of LRDs can be explained via a viewing angle-dependent nuclear structure in which an optically thick, clumpy gaseous torus surrounds the central accretion disk, with broad-line clouds and absorbers distributed along the less-obscured polar directions.
\end{abstract}


\section{Introduction}

The discovery of little red dots (LRDs; \citealt{Matthee2024_lrd_wfss}) with JWST has revealed a previously unrecognized population of sources at high redshift. LRDs are optically compact objects with a unique V-shaped spectral energy distribution (SED) that is blue in the rest-frame UV and red in the optical. Follow-up spectroscopic observations indicate that the majority of LRDs have broad emission lines with full width at half maximum (FWHM) $\gtrsim 1000\ {\rm km~s^{-1}}$ \citep{Greene2024_lrd_spec, Hviding2025}, a feature commonly observed in local active galactic nuclei (AGNs). In addition, the extreme compactness ($r_{e}\lesssim 100\ {\rm pc}$) \citep{Kokorev2024_lrd, Akins2025, Kocevski2025_lrd} and tentative variability observed in individual sources \citep{Zhang2025_lrd_var, Zhang2026_lrd_lense_var} add to the evidence that LRDs are most likely AGNs in the high-redshift Universe.

Compared to the well-established scaling relation between black hole (BH) mass and galaxy mass in the local Universe \citep{Magorrian1998, Kormendy2013, Greene2020}, many LRDs have overmassive BHs, as indicated by a wide range of independent approaches. Constraints from imaging-based structural modeling \citep{Chen2025}, joint SED and spectral decompositions \citep{Labbe2025_lrd, Maiolino2025}, and large-scale environment statistics \citep{Matthee2025_lrd_env} all converge on stellar masses that lie far below those implied by local BH--galaxy relations, frequently by more than 2 orders of magnitude. Most strikingly, in at least one system, spectro-astrometric measurements of the rotation velocity require the stellar mass to be smaller than the BH mass itself, consistent with the presence of a nearly ``naked'' BH \citep{Juodzbalis2025_bh_mass}. As they are commonly seen only at $z\gtrsim 4$ \citep{Greene2024_lrd_spec, Kokorev2024_lrd, Ma2025}, LRDs presumably represent the initial phase of BH mass assembly \citep[see also][]{Inayoshi2025_lrd_duty_cycle}, where the surrounding stellar structures have not yet come into being.

LRDs exhibit panchromatic SEDs and spectral signatures that are markedly different from those of local type~1 AGNs \citep[see the review by][]{Inayoshi_ho_lrd_rev}. A significant fraction ($10\%-20\%$) of LRDs show Balmer absorption features superposed on the emission lines \citep{Matthee2024_lrd_wfss, Zhuang2025_lrd, Lin2025_hae}, a phenomenon that is extremely rare in local AGNs \citep{Shangguan2026}. Spectral breaks near the Balmer limit at rest-frame $3646\ {\rm \AA}$ are also frequently observed among LRDs. Despite their resemblance to the Balmer breaks seen in relatively evolved stellar populations, the exact shape of the Balmer break in LRDs cannot be reproduced solely by stellar population synthesis models \citep{Ma2025_bb_shape}. Besides, the strength of the Balmer break in some LRDs also far exceeds the maximum value that stellar populations can generate \citep{Wang2024, Naidu2025_bh_star}, suggesting a distinctive physical origin.

Even more substantial differences between high-redshift LRDs and low-redshift AGNs exist in terms of the continuum shape. Low-redshift quasars show spectral uniformity, featuring a broken power-law continuum with a power-law index ($f_{\nu}\propto \nu^{\alpha_{\nu}}$) of $\alpha_{\nu, {\rm UV}}=-0.44$ from $1300$--$5000\ {\rm \AA}$ and $\alpha_{\nu, {\rm opt}}=-2.45$ redward of $5000\ {\rm \AA}$ \citep{VandenBerk2001}, both with an uncertainty of $\sim 0.1$ in the spectral index. For LRDs, while the UV continuum slope is generally consistent with that of typical quasars, the optical continuum is much redder. Despite the selection criterion of $\alpha_{\nu, {\rm opt}}<-2$, most LRDs have $\alpha_{\nu, {\rm opt}}<-2.5$, including some extreme cases with $\alpha_{\nu, {\rm opt}}\approx -5$ \citep{Kocevski2025_lrd}. Initially, the redness was assumed to originate from reddening, where an intrinsically blue quasar spectrum is attenuated by dust with $A_{V}=3-5$ mag. Indeed, many LRDs show a large Balmer decrement between H$\alpha$ and H$\beta$ \citep{Killi2024, Nikopoulos2025}. The ratios between the broad components of various higher-order Balmer lines are also consistent with the prediction from large dust attenuation of $A_{V}\gtrsim 3$ mag in the broad-line region (BLR) \citep{Nikopoulos2025}. However, the absence of dust re-emission from the rest-frame NIR to FIR in LRDs indicates a lack of both hot ($T\gtrsim 500\ {\rm K}$) and cold ($T\lesssim 75\ {\rm K}$) dust grains \citep{PerezGonzalez2024, Akins2025, Setton2025, Wang2025}. Considering a wide range of dust extinction laws and spatial distributions, \citet{Chen2025_lrd_dust} state that LRDs on average should have $A_V\lesssim 1.0-1.5$ mag, indicating that the observed redness is likely intrinsic instead of arising from dust reddening. The lack of dust emission is also consistent with the low metallicities measured from emission-line diagnostics \citep{Ivey2026, Korber2026, Maiolino2026_metallicity} or nebular continuum shape \citep{Chen2025_extended_comp}, as well as the undermassive or even absent host galaxies in LRDs \citep{Furtak2024, Chen2025, Jones2026, Juodzbalis2025_bh_mass}. Collectively, the spectral and continuum discrepancies point to a nuclear structure in LRDs that may depart from the conventional AGN paradigm \citep{Antonucci1993, Netzer2015}.

A commonly adopted approach to resolving the aforementioned discrepancies is to invoke a dense gaseous medium surrounding the central engine \citep{Inayoshi_ho_lrd_rev}. The gaseous medium has sufficiently high density ($n_{\rm H} \gtrsim 10^9\ {\rm cm^{-3}}$) to allow collisional excitation of hydrogen to higher energy levels. The medium is also mostly neutral, so Ly$\alpha$ photons are trapped because of the large optical depth. Both effects result in an elevated population of hydrogen atoms at the $n=2$ energy level, making the medium no longer optically thin to any transition toward the $n=2$ state, thus naturally generating the prominent Balmer break and Balmer absorption lines. The high-density gas can also potentially explain the large broad-line Balmer decrement. Under Case~B conditions, the ratio between H$\alpha$ and H$\beta$ emissivities is 2.86 at $10,000\ {\rm K}$ \citep{Draine2011}. If the medium remains mostly optically thin to Balmer transitions, the intrinsic line flux ratio should equal the emissivity ratio, since most photons can escape the medium, and any deviation from this ratio is attributed to dust attenuation.

However, Balmer lines gradually become optically thick in a high-density medium. The optical depths of the \ha\ and \hb\ lines change differently as a function of density, resulting in different values of the Balmer decrement, which can be as large as 20 without invoking dust \citep{Capriotti1964, Netzer1975, Kwan1981, Korista1997, Yan2025}. The high-density gaseous medium also gives a consistent explanation of the optical-to-NIR continuum shape. With a semi-analytical model describing the structure of the accretion flow, radiative-transfer calculations from \citet{Liu2025_sphere} show that the gas configuration would result in a blackbody-like spectrum with an effective temperature close to $5000\ {\rm K}$, which can match both the red optical continuum and the IR flattening in LRDs. The model prediction for the continuum shape was later confirmed in three local LRD analogs, for which the infrared continuum can be accurately measured out to rest-frame $2\ {\rm \mu m}$ \citep{Lin2025_local_lrd}.

Despite its successes, the dense gaseous medium scenario faces several conceptual challenges. In order to reproduce the observed blackbody-like optical continuum, the medium must be highly optically thick. However, if both the accretion disk and the BLR are embedded within such a medium, photons with energies above the Balmer limit in the UV continuum, as well as broad Balmer emission lines, would be strongly absorbed and unable to escape, in tension with observations. Several studies have attempted to alleviate this issue by invoking non-spherical geometries for the dense gas, in which both the broad emission lines and the UV continuum from the accretion disk can escape along lines of sight with lower covering fractions \citep{Lin2025_local_lrd, Matthee2026_lrd_prof, Sneppen2026}. Alternatively, some models retain a fully spherical dense medium but introduce an additional compact host-galaxy component exterior to the medium to account for the observed UV continuum. In the so-called ``black hole star'' scenario, broad Balmer line photons produced within the dense medium undergo repeated absorption and re-emission by hydrogen atoms with an enhanced population in the $n=2$ level before escaping from the surface \citep{deGraaff_bh_star, Naidu2025_bh_star, Sun2026}. This process is commonly referred to as resonant scattering \citep{Chang2025}. A related but distinct framework proposed by \citet{Pang2026} considers broad-line clouds embedded within a dense, clumpy gaseous medium to reproduce the observed optical continuum shape. In another model, \citet{Inayoshi2026_ha_pos} suggest that gas located at the outer boundary of a spherical dense medium, photoionized by stellar radiation from the host galaxy, can reach dynamical velocities of $\sim 1000\ {\rm km\ s^{-1}}$, thereby forming the BLR. Overall, the detailed structure of the nuclear gaseous environment remains poorly constrained. The relative spatial configuration of the dense medium, broad-line clouds, accretion disk, and potential Balmer line absorbers is still under active debate, and the origin of the observed UV continuum remains uncertain. 

In this work, we carefully examine the Balmer line properties and continuum shapes of 14 LRDs using prism and medium-resolution spectra from NIRSpec/MSA. By studying the correlations among the continuum, broad and narrow emission lines, and absorption features, we aim to obtain a comprehensive understanding of the structure of different nuclear components, as well as how they correspond to the observed spectral features. In Section~\ref{data_sample}, we introduce the public imaging and spectroscopic data adopted in this work, as well as our sample selection procedures. Section~\ref{spec_fitting_sec} discusses the emission-line modeling techniques, and the results are presented and discussed in Sections~\ref{results} and \ref{discussion}. Throughout the paper, we assume a flat cosmology with $\Omega_{m}=0.3111$, $\Omega_{\Lambda}=0.6889$, and $H_0=67.66\ {\rm km}\ {\rm s^{-1}}\ {\rm Mpc^{-1}}$ \citep{Planck2018}.

\section{Data and Sample}\label{data_sample}

We broadly follow the methodology of \citet{Zhang2025_nrl_lrd} for the LRD sample selection. LRDs are identified based on the rest-frame UV-optical continuum slope and compactness in the rest-frame optical. Most public JWST surveys have NIRCam imaging from 1--4.4~$\mu$m, covering the rest-frame optical SED for sources at $2\lesssim z\lesssim 7$. However, at $z\lesssim 6.4$, the rest-frame UV continuum begins to shift out of the NIRCam wavelengh coverage, requiring HST imaging data to measure the UV continuum slope. To achieve self-consistent photometric selection, prior to obtaining photometry using \cite{Kron1980} apertures, we reduce raw JWST images to a pixel scale and World Coordinate System (WCS) frame consistent with those of archival HST images. NIRSpec/Prism and medium-resolution spectra are obtained from the Dawn JWST Archive (DJA).\footnote{https://dawn-cph.github.io/dja/} The final LRD sample is constructed by imposing constraints on the continuum slope and source compactness in the F444W band. A concise overview of the data reduction and sample selection procedures is provided in the following sections. 

\subsection{Photometric and Spectrosopic Data}

Our selection covers several widely studied deep fields, including the A2744, CEERS, GOODS-S, GOODS-N, COSMOS, and UDS fields. We reduce and drizzle all the publicly available imaging data from stage 1, using a combination of the {\tt\string JWST Calibration Pipeline} (v1.12.5), the CEERS NIRCam imaging reduction procedures \citep{Bagley2023}, and additional customized routines \citep{Zhang2025_lrd_var,Zhang2025_nrl_lrd}.
All NIRCam mosaics are produced with a pixel scale of $0\farcs03$ and are registered to archival HST images from the CANDELS and 3D-HST surveys, with the WCS coordinates from the HST images taken as references. For multi-band photometry, all images are matched to the point-spread function of the F444W band, and source detection is performed on the F444W image using {\tt\string SExtractor} \citep{Bertin_sextractor} in a two-step “cold + hot” mode. 
Initial photometry is measured within a small elliptical aperture defined by 1.1 times the Kron radius ($k=1.1$), with the minimum allowed Kron radius set to $R_{\rm min}=1.6$ pixels. The flux is subsequently corrected to match the default Kron aperture ($k = 2.5$, $R_{\rm min}=3.5$) in each band to recover total fluxes consistently across filters.

Spectroscopic data are obtained from DJA release version 4.4. Detailed descriptions of the data-reduction procedures are provided in \citet{Heintz2024} and \citet{deGraaff_rubies2025}. In brief, the uncalibrated exposures for each NIRSpec/MSA observation are processed with the {\tt\string Detector1Pipeline} and {\tt\string Spec2Pipeline} modules of the standard JWST pipeline, yielding flat-fielded and flux-calibrated two-dimensional spectra for each target. The {\tt\string msaexp} package is then used to perform background subtraction, apply WCS calibration, and combine multiple dithered exposures onto a rectified pixel grid. Slit-loss corrections are computed by assuming an idealized point-source morphology, projecting the emission onto the detector plane, and subsequently extracting one-dimensional spectra. The wavelength-dependent line-spread function (LSF) is derived simultaneously and has been shown to provide a more accurate characterization than the pre-launch instrumental resolution reported in the JWST documentation (JDox) \citep{deGraaff_lsf2024}. For targets observed with NIRSpec/Prism, spectroscopic redshifts are estimated with {\tt\string msaexp} by fitting the {\tt\string agn\_blue\_sfhz\_13} template used in {\tt\string EAZY} \citep{Brammer2008_eazy} to the prism spectra.

\subsection{LRD Sample Selection}

LRDs are selected from the DJA NIRSpec/MSA spectroscopic sample following the procedures described in \citet{Zhang2025_nrl_lrd}. We first identify LRD candidates based on their characteristic V-shaped SEDs, quantified by the UV and optical continuum slopes following criteria proposed by \citet{Kocevski2025_lrd}. For all sources at $z>2$ with NIRSpec/Prism observations, the UV continuum slope $\alpha_{\rm UV}$ is derived using photometric bands with rest-frame pivot wavelengths $\lambda_{\rm pivot,rest}\in[1350,3645]\ {\rm \AA}$, while the optical continuum slope $\alpha_{\rm opt}$ is measured using bands with $\lambda_{\rm pivot,rest}\in[3300,8000]\ {\rm \AA}$. Contributions from strong emission lines, including H$\beta$, [O III] $\lambda\lambda4959,5007$, and H$\alpha$, are modeled by convolving the prism spectra with the transmission curves of the corresponding filters and are subsequently subtracted from the broadband photometry prior to continuum fitting. Sources with $\alpha_{\rm UV}>-1.63$ and $\alpha_{\rm opt}<-2$ are classified as LRD candidates. In this step, we identify 247 candidates out of 7,645 sources with usable NIRSpec/Prism observations at $z>2$.

An additional compactness requirement is imposed on the half-light radius ($R_{\rm F444W}$) measured with {\tt\string SExtractor} on the F444W image. We first derive a relation between $R_{\rm F444W}$ and apparant magnitude for field stars \citep{Zhang2025_nrl_lrd}. A source is classified as compact if its $R_{\rm F444W}$ is smaller than 1.5 times the $R_{\rm F444W}$ of stars with similar magnitude. In total, 126 LRDs satisfy the selection criteria.

To enable reliable decomposition of the Balmer emission-line profiles, additional constraints are imposed on the spectral resolution and SNR of the available observations. Specifically, ${\rm H}\alpha$, ${\rm H}\beta$, ${\rm H}\gamma$, and \OIII\ must all be covered by NIRSpec/MSA observations obtained with the medium-resolution (G140M, G235M, G395M) gratings. The presence of [\ion{O}{3}] is required to constrain the narrow-line profile. In addition, a SNR threshold of 10 is applied within $\pm 10{,}000\ {\rm km\ s^{-1}}$ relative to \ha, and within $\pm 5000\ {\rm km\ s^{-1}}$ relative to [\ion{O}{3}], as \ha\ is primarily used to model the potential broad emission component and [\ion{O}{3}] for constraining the narrow-line profile. 

Although high-resolution spectra are available for some sources, we restricted the analysis to the medium-resolution data because their velocity resolution of $\rm FWHM \approx 300\ {\rm km\ s^{-1}}$ can resolve the typically broad emission lines in LRDs, whose line widths are generally $\gtrsim 1000\ {\rm km\ s^{-1}}$. Moreover, adopting uniform medium-resolution data preserves a relatively large sample while ensuring consistent spectral quality and full coverage of all four key emission lines. In total, 11 LRDs satisfy all selection criteria. When multiple medium-resolution spectra cover the same emission line, the observation with the highest SNR is used in the analysis. Besides this primary sample, we also included SID-28074, SID-42046, and SID-55604 mainly for the analysis of the velocity-resolved Balmer decrement (Section~\ref{stratified_gas}). The three additional sources satisfy our LRD selection criteria. Their medium-resolution spectra only cover H$\alpha$ and H$\beta$, but the high SNR makes them ideal to derive the Balmer decrement profiles. The three sources also have prominent Balmer absorption features, which will be discussed in Section~\ref{tau_ratio}. Our final sample consists of 14 LRDs (Table~\ref{results_table}).

\section{Spectral Fits}\label{spec_fitting_sec}

This section describes the models and procedures used to measure the Balmer line profiles and Balmer break strength. Emission-line fitting is performed with a Python-based, custom package {\tt\string GalSpec}\footnote{\url{https://github.com/jyshangguan/GalSpec}} \citep{Kuhn2024, Santos2025, Shangguan2026, Li2025_sagan_mcmc}, which provides Gaussian emission-line and absorption-line models compatible with the {\tt\string Astropy.modeling} package. The line models are parameterized in velocity space (line width and centroid shift), to facilitate tying different parameters among different transitions. Best-fit parameters and associated uncertainties are obtained through MCMC sampling using {\tt\string emcee}. 

The Bayesian information criterion (BIC; \citealt{Schwarz1978}) is used as an important, although not exclusive, metric for comparing the quality of different models. It is defined as

\begin{equation}
{\rm BIC}=k\ln N+\chi^2 ,
\end{equation}

\noindent
where $k$ is the number of free parameters and $N$ is the number of data points. A smaller BIC value indicates better model performance, and a difference of more than 10 between two BIC values is commonly interpreted as strong evidence favoring the model with the lower BIC \citep[e.g.,][]{Jeffery1961, Burnham2002}.

	\subsection{Spectral Line Models}\label{spec_line_model}

Broad lines with FWHM $\gtrsim 1000\ {\rm km\ s^{-1}}$ in \ha, \hb, and \paa\ are observed in a large fraction of LRDs \citep[e.g.,][]{Greene2024_lrd_spec, Matthee2024_lrd_wfss, Lin2025_local_lrd, Naidu2025_bh_star}, and $10\%-20\%$ show absorption troughs superimposed on the emission lines \citep{Matthee2024_lrd_wfss, Zhuang2025_lrd,  Lin2025_hae}, indicating high-density absorbers along the line-of-sight \citep{Inayoshi2025_bd_ba}. Even in narrow-line LRDs, whose broad wings are statistically insignificant, ${\rm H}\alpha$ often appears slightly broader than the narrow forbidden line [\ion{O}{3}] \citep{Zhang2025_nrl_lrd}. By analogy with low-redshift type 1 AGNs, the broad component is commonly interpreted as tracing gas motions in the immediate vicinity of the central BH. Dynamically broadened profiles are typically modeled using one or more Gaussian functions \citep[e.g.,][]{Kollatschny2013, Maiolino2025, Juodzbalis2025}.

We fit the emission lines in the medium-resolution spectra using a combination of multiple Gaussians. Our model includes up to five components: a single Gaussian for the narrow-line emission, one or two Gaussians for the broad-line emission, a first-order polynomial for the underlying continuum, and, when required, an absorption component. The absorption component is described following the formalism of \citet{DEugenio_2025}:

\begin{equation}\label{eq_rabs}
\begin{aligned}
& R_{\rm abs}\equiv f/f_0=1-C_f+C_f\exp\left[-\tau(v)\right], \\
& \tau(v)=\tau_0\exp\left[-\frac{(v-v_0)^2}{2\sigma_{v}^2}\right],
\end{aligned}
\end{equation}

\noindent
where $f_0$ and $f$ are the fluxes before and after absorption, respectively, $C_f$ the covering factor of the absorbing medium, $\tau_0$ the optical depth at line center, and $v_0$ and $\sigma_{v}$ the systemic velocity shift and velocity dispersion of the absorber.

\subsection{Determination of the Narrow-line Profile}\label{o3_fit}

To decompose the narrow and broad components of the Balmer lines, we first examine the forbidden lines to determine the narrow-line profile, which is expected to be largely limited by the instrumental spectral resolution. We focus on \OIII, which is the most prominent forbidden line in LRD spectra. The fit is performed over the spectral region within $\pm 6000\ {\rm km\ s^{-1}}$ of [\ion{O}{3}] $\lambda 5007$, using a single Gaussian for each line of the doublet and a first-order polynomial for the underlying continuum. The FWHM and systemic velocity of the doublet are tied, with the amplitude ratio fixed to the theoretical value of 2.98 \citep{Osterbrock2006}. For each source, running 6000 MCMC steps suffices for convergence. In 13 of the 14 LRDs, the best-fit FWHM of [\ion{O}{3}] is smaller than that of the theoretical LSF calculated by \citet{deGraaff_lsf2024}. The only exception is SID-47509, for which the FWHM exceeds the LSF by only $6\ {\rm km\ s^{-1}}$. Since [\ion{O}{3}] is unresolved or close to the spectral resolution limit in all sources, we only use a single Gaussian in the fit. The discrepancy between the FWHM measured from the unresolved [\ion{O}{3}] line and the theoretical LSF exposes the limitations of the LSF model. For sources with unresolved [\ion{O}{3}], we therefore adopt the best-fit [\ion{O}{3}] model as the effective LSF.

\begin{figure*}
	\centering
	\includegraphics[width=1\textwidth]{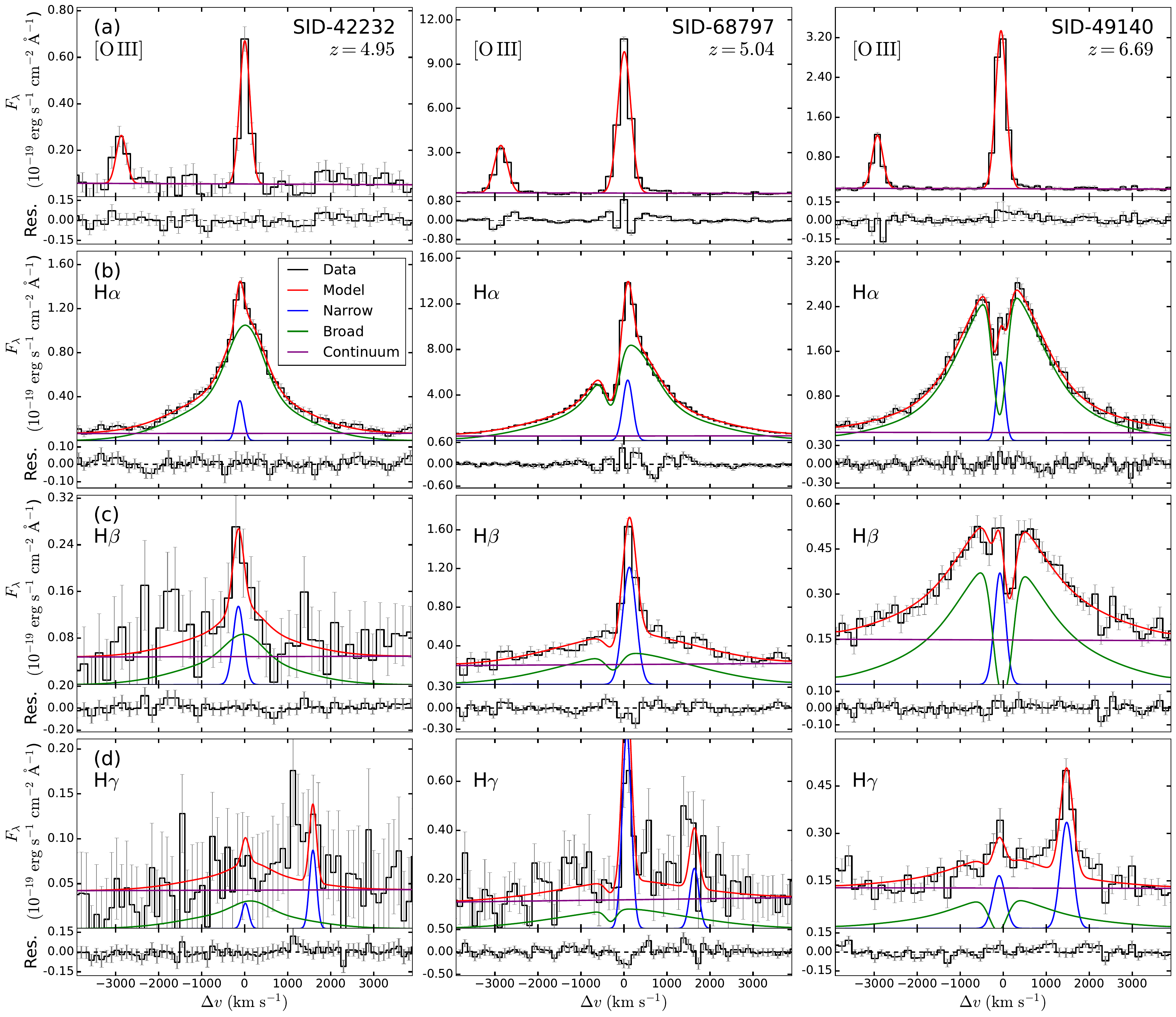}
	\caption{Spectral fits for (a) [\ion{O}{3}] $\lambda\lambda 4959,5007$, (b) H$\alpha$, (c) H$\beta$, and (d) H$\gamma$ for SID-42232 (left), SID-68797 (middle), and SID-49140 (right). The top row in each panel shows the data and uncertainty in black line with gray errorbars, and the continuum, narrow-line component, broad-line component, and total model purple, blue, green, and red lines, respectively. Residuals are displayed in the bottom row. }
	\label{spec_fitting}

\end{figure*}

\begin{figure*}
	\centering
	\includegraphics[width=1\textwidth]{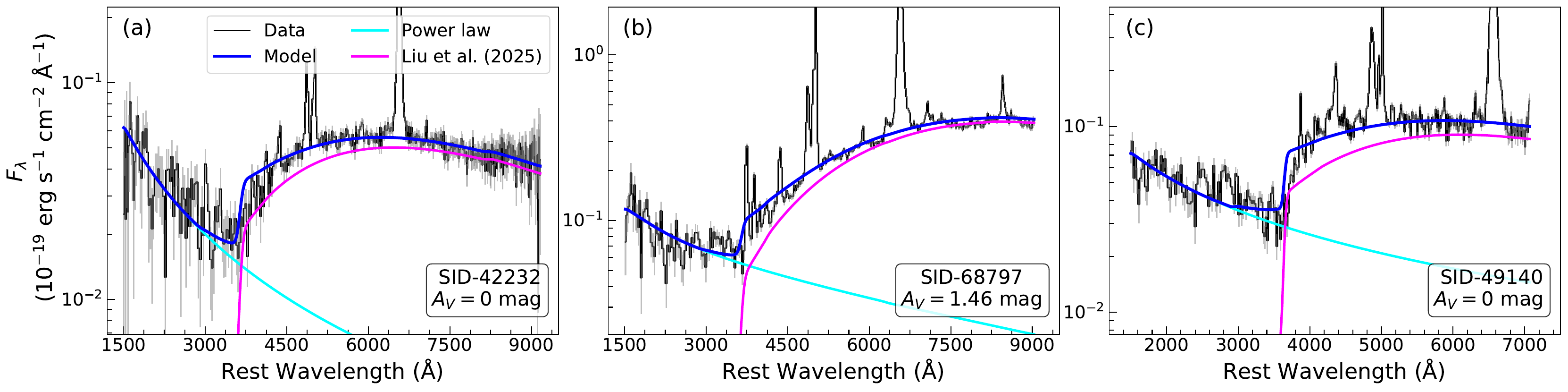}
	\caption{Continuum fits of the prism spectra for (a) SID-42232, (b) SID-68797, and (c) SID-49140. The data are plotted as black solid lines, with the uncertainties shown in gray. The models are shown by the blue solid lines, with the power-law component and the spherical dense gas model of \citet{Liu2025_sphere} plotted in cyan and magenta, respectively. The adopted values of dust extinction are labeled in the plot. }
	\label{continuum_spec_fitting}
\end{figure*}

\subsection{Decomposition of the Balmer Lines}\label{Balmer}

The Balmer-line profiles generally consist of a broad component, produced by ionized gas presumably in the vicinity of the massive BH, and a narrow component emitted on larger scales. The decomposition becomes more complicated when absorption features are present. We therefore follow techniques widely adopted in AGN spectral decomposition, in which the narrow Balmer-line profile is fixed to that of the forbidden lines under the assumption that they arise from a similar region and share similar kinematics (e.g., \citealt{Ho1997,HoKim2009,Shangguan2026}). Because H$\alpha$ typically has the highest SNR among the Balmer lines, we first fit its profile to determine the number of Gaussian components required to describe the broad-line emission. We then fit the H$\beta$ and H$\gamma$ profiles by modeling their broad components with the same number of Gaussian components as used for H$\alpha$, while tying the line widths and velocity offsets to those of the corresponding H$\alpha$ components. This approach reduces the number of free parameters while retaining sufficient flexibility to account for profile differences among the Balmer lines.

The profile of [\ion{O}{3}] constrains the shape of the narrow component of the Balmer lines. We cannot constrain the intrinsic FWHM of the narrow lines for the 13 sources with unresolved [\ion{O}{3}]. We therefore treat the narrow component of the Balmer lines \ as unresolved, estimating the effective LSF from the best-fit [\ion{O}{3}] model. Adopting the wavelength dependence of the theoretical LSF, we estimate the FWHM of narrow \ha\ as

\begin{equation}
{\rm FWHM}_{\rm H\alpha, n}
=
{\rm FWHM}_{\rm [O\,III]}
\times
\frac{{\rm FWHM}_{\rm {H\alpha}, R}}
{{\rm FWHM}_{\rm {[O\,III]}, R}} .
\end{equation}

\noindent
Here, ${\rm FWHM}_{\rm H\alpha, n}$ is the FWHM of the narrow \ha\ component, ${\rm FWHM}_{\rm H\alpha, R}$ and ${\rm FWHM}_{\rm [O\,III],R}$ are the FWHM of the theoretical LSF at the wavelength of \ha\ and [\ion{O}{3}] $\lambda5007$, respectively, and ${\rm FWHM}_{\rm [O\,III]}$ is the best-fit FWHM of [\ion{O}{3}]. For the one LRD with resolved [\ion{O}{3}]  emission, we derive the FWHM of narrow \ha\ through

\begin{equation}
\small
{\rm FWHM}_{\rm H\alpha, n}
=
\sqrt{
{\rm FWHM}_{\rm [O\,III]}^2
-
{\rm FWHM}_{\rm [O\,III], R}^2
+
{\rm FWHM}_{\rm H\alpha, R}^2
} .
\end{equation}

\noindent
The same procedure follows for H$\beta$ and H$\gamma$.

\subsubsection{Decomposition of ${\rm H}\alpha$}\label{Halpha}

The profiles of the broad Balmer lines are generally similar in an individual AGN \citep{Osterbrock1982, Shuder1982, Ilic2012}, allowing the high-SNR \ha\ line to constrain the model components used for the weaker \hb\ and \hg\ lines. We fit \ha\ with five independent model sets: a single narrow component; a narrow component plus an additional broad component; a narrow plus a broad component with absorption; a narrow plus two broad components; and a narrow plus two broad components with absorption. A first-order polynomial is included to describe the underlying continuum. The broad components are required to have FWHMs larger than that of the narrow component. For each model, we perform MCMC sampling with 10,000 steps to ensure convergence.

We first test whether a broad component is required by comparing the pure narrow-line model with the narrow$+$broad-line model. A broad component is retained only if its inclusion lowers the BIC by more than 10 and if its SNR, defined as the best-fit flux divided by its uncertainty, exceeds 3. We then apply the same criteria to determine whether a second broad component is needed. Finally, we test for Balmer absorption by comparing the best-fit models with and without an absorber. The absorption component is considered significant only if its inclusion lowers the BIC by more than 10.

When an absorption component is included, the model is written as

\begin{equation}
f_{m}=f_{n}+(f_{b}+f_{\rm cont})\cdot R_{\rm abs},
\end{equation}

\noindent
where $f_{m}$, $f_{n}$, $f_{b}$, and $f_{\rm cont}$ denote the total model flux, the narrow-line component, the broad-line component, and the continuum, respectively. The broad-line component $f_{b}$ includes either one or two Gaussians. This formalism assumes that the absorber is located outside the BLR and the optical continuum-emitting region, so that absorption affects only the broad-line emission and the continuum, following \citet{Brazzini2025} and \citet{DEugenio_2025}. As shown in Section~\ref{results}, the velocity dispersions of the absorbers are larger than those of the unresolved narrow emission lines, supporting our choice to exclude the narrow-line component from the absorption term.

\begin{figure*}
	\centering
	\includegraphics[width=0.95\textwidth]{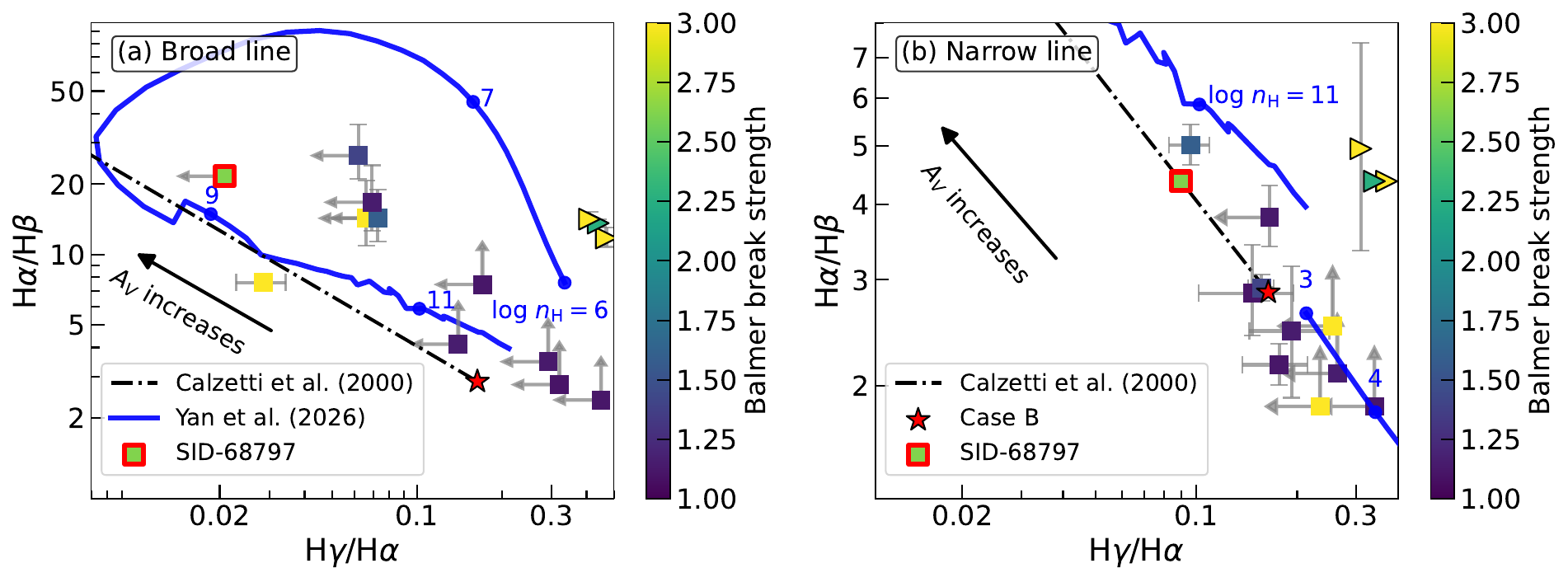}
	\caption{The H$\alpha$/H$\beta$ and H$\beta$/H$\gamma$ line ratio for the (a) broad and (b) narrow line. The squares with errorbars are measurements for our sample. Since SID-28074, SID-42046, and SID-55604 do not have medium-resolution spectra coverage of the \hg\ line, we plot the H$\alpha$/H$\beta$ ratio as triangles with errorbars on the right side of each panel, with the horizontal positions slightly shifted for clarity. All data points are color-coded by the Balmer break strength. The blue solid lines show the theoretical predictions of \citet{Yan2025} for $N_{\rm H}=10^{23}\ {\rm cm^{-2}}$ and ${\rm log}\ U=-1.5$, with blue dots denoting $n_{\rm H}=(10^6,\ 10^7,\ 10^9,\ 10^{11})\ {\rm cm^{-3}}$ in panel (a) and $n_{\rm H}=(10^3,\ 10^4,\ 10^{11})\ {\rm cm^{-3}}$ in panel (b). The standard Case~B value of H$\alpha$/H$\beta=2.86$ and H$\gamma$/H$\alpha=0.16$ \citep{Draine2011} is plotted as a red star. The dash-dotted line represents the Case~B value attenuated by the curve of \citet{Calzetti2000}, with $R_{V}=4.05$. For clarity, we shifted slightly the point for SID-50716, whose actual values are given in Table~\ref{results_table}.
	 }
	\label{line_ratio_bb}
\end{figure*}

\subsubsection{Decomposition of ${\rm H}\beta$ and ${\rm H}\gamma$}\label{hbg}

Although the profiles of different Balmer transitions are not necessarily identical \citep{Chang2025,Matthee2026_lrd_prof}, the limited SNR of \hb\ and \hg\ in our LRD sample preclude independent determinations of their line profiles. We therefore fit all three Balmer lines simultaneously to improve the overall constraints. In this approach, the high-SNR \ha\ line anchors the profile shape, while the \hb\ and \hg\ data primarily constrain the relative fluxes. We adopt the same number of Gaussian components for \hb\ and \hg\ as for \ha, and tie their FWHMs and relative velocity offsets to those of the corresponding \ha\ components. The amplitudes of the Gaussian components are allowed to vary independently among the three Balmer lines, providing flexibility for differences in their relative component strengths. The absolute velocity shift is left as a free parameter to account for possible wavelength-calibration uncertainties between spectra obtained from different programs. Whenever absorption is included, we tie the FWHM, velocity offset, and covering factor of the absorption component across the three Balmer lines.

The two parameters describing the absorption profile, the covering factor $C_f$ and line-center optical depth $\tau_0$, are partially degenerate. When $\tau_0 \ll 1$, Equation~(\ref{eq_rabs}) reduces to $R_{\rm abs} \approx 1 - C_f \cdot \tau(v)$. In this optically thin limit, the absorption trough follows the Gaussian profile of $\tau(v)$, and $\tau_0$ and $C_f$ are fully degenerate. When $\tau_0 \gg 1$, $R_{\rm abs}$ approaches $1-C_f$ near the line center, producing a flat-bottomed absorption trough whose depth is set by $C_f$. At low spectral resolution, this flat core may not be resolved and can mimic a narrower, Gaussian-like absorption profile. In this case, a decrease in $C_f$ is degenerate with a decrease in $\tau_0$, as both make the absorption trough shallower. 

However, since Balmer absorption lines originate from the same lower energy level, the optical-depth ratios between different Balmer lines are primarily determined by the oscillator strengths, provided that the lines arise from the same absorber and share the same covering factor. These ratios can be derived from basic atomic physics\footnote{Einstein $A$ coefficients for the transitions are taken from the NIST Atomic Spectra Database: https://www.nist.gov/pml/atomic-spectra-database.}, giving $\tau_{0,{\rm H}\alpha}/\tau_{0,{\rm H}\beta}=7.13$ and $\tau_{0,{\rm H}\alpha}/\tau_{0,{\rm H}\gamma}=21.18$, unless the relative $2s/2p$ populations deviate from their statistical-weight ratio. Ly$\alpha$ pumping is the most relevant process for producing the $n=2$ H population required for Balmer absorption \citep{Hall2007}; however, it increases these ratios by only $\lesssim 10\%$. In our simultaneous fitting of the Balmer lines in LRDs, the \ha\ optical depth is typically $\gg 1$. The value of $\tau_{0,{\rm H}\alpha}$ can still be constrained as long as the optical depth of \hb\ or \hg\ is $\lesssim 1$, because their $\tau_0$ values are tied. Otherwise, we can only derive a lower limit on $\tau_{0,{\rm H}\alpha}$. Nevertheless, we can robustly determine $C_f$, as it reflects the relative absorption-trough depths across the three Balmer lines \citep{Shangguan2026}.

To better constrain the optical depth, covering factor, and the velocity dispersion $\sigma_{v}$ of the absorber, we follow the procedure of \citet{Shangguan2026} for modeling LRDs with absorption lines.
Specifically, we convolve the intrinsic model with the LSF as

\begin{equation}
f_{\rm conv}(\lambda)=\left[ \left( f_{\rm cont}(\lambda)+f_{\rm broad}(\lambda) \right) f_{\rm abs}(\lambda) \right] \otimes G_{\rm LSF}(\lambda),
\end{equation}

\noindent
where $f_{\rm cont}$ and $f_{\rm broad}$ are the continuum and the total broad line models, and $G_{\rm LSF}$ is the wavelength-dependent LSF. For simplicity, we do not convolve the narrow-line component; instead, we add it to the convolved model during the fitting, because the narrow lines are mostly unresolved. During fitting, the optical depth ratios are fixed to their theoretical values, and the covering factor is tied across all lines. \citet{Shangguan2026} applied this fitting method to low-redshift AGNs with Balmer absorption lines in their spectra, which have spectral resolutions of $\sim$2000--5000. They found that convolving the model with the LSF, together with tying the covering factor and optical depth ratios, allows reliable recovery of covering factors from $0.1$ to $1$ and \ha\ optical depths up to $\sim 100$. As our NIRSpec medium-resolution spectra have a lower resolution of $R\approx 1000$, a similar range of parameter priors ($C_f = 0-1$ and $\tau_{\rm H\alpha} = 0.01-100$) is adequate for our purposes.

The complexity of our model, which includes absorption lines, compels us to increase the MCMC sampling to $10^5$ steps to ensure that the simultaneous fitting of \ha, \hb, and \hg\ converges. Figure~\ref{spec_fitting} illustrates the simultaneous line fitting results for three LRDs, representing a typical case for sources without absorption (SID-42232), sources with blueshifted absorption (SID-68797), and sources with a relatively centered absorption (SID-49140). We obtain the posterior distribution for each parameter and calculate the 16th, 50th, and 84th percentile to estimate the median value and the upper and lower $1\, \sigma$ uncertainties. Following \citet{Shangguan2026}, a measurement is considered an upper (lower) limit when the median value lies within $3\,\sigma$ of the lower (upper) boundary of the posterior distribution. In such cases, the 0.3rd and 99.7th percentiles of the posterior distribution are taken as the lower and upper limits ($3\,\sigma$), respectively. Otherwise, the median value of the posterior is reported, and the differences between the 16th, 50th, and 84th percentiles are quoted as the lower and upper $1\,\sigma$ uncertainties.

Three sources, SID-28074, SID-42046 and SID-68797, show large narrow-line Balmer decrements of ${\rm H}\alpha/{\rm H}\beta \gtrsim 8$ when the narrow Balmer-line components are fitted freely. Taken at face value, this Balmer decrement corresponds to an extinction of $A_{V} \gtrsim 2.5$ mag, assuming the attenuation curve of \citet{Calzetti2000}. 
We consider the high narrow-line Balmer decrements likely unreliable, because the Balmer lines, especially H$\alpha$, show smooth profiles in which the broad and narrow components are difficult to separate. The absorption troughs further complicate a robust decomposition. In addition, the high dust attenuation level is also in contradiction with the values predicted by \citet{Chen2025_lrd_dust} based on the mid-IR photometry and morphology of four typical LRDs, which should be less than 2.5 even in the most conservative case assuming a flat Orion Nebular attenuation curve \citep{Baldwin1991}. 
We therefore fix the Balmer decrement to a typical AGN value and test whether a reasonable solution can be obtained. Specifically, we set ${\rm H}\alpha/{\rm H}\beta=4.37$, the median observed value for low-redshift AGNs \citep{Lu2019}. This value is higher than the Case~B recombination value, suggestive of moderate dust attenuation ($A_V=1.46$). To maintain consistent attenuation following the \citet{Calzetti2000} curve, we adopt ${\rm H}\gamma/{\rm H}\beta=0.09$. To further assess how model assumptions affect narrow-line strengths, we also fit the line profiles with exponential broadening and a P-Cygni absorption profile for the five sources exhibiting blue-shifted absorption troughs. As summarized in Appendix~\ref{exp_pcygni}, the derived narrow Balmer line fluxes, and hence the decrements, depend on the adopted model, though a detailed exploration of this dependence is beyond the scope of this work. Both methods nonetheless indicate moderately high Balmer decrements for these sources.

\subsection{Continuum Measurements}\label{cont_fitting}


We first measure directly from the prism observations the strength of the Balmer break, defined as the ratio of the flux density averaged over $\lambda_{\rm rest} \in [3000, 3500]\ {\rm \AA}$ to that averaged over $\lambda_{\rm rest} \in [4000, 4500]\ {\rm \AA}$, with the flux density expressed per unit frequency, as in previous studies \citep[e.g.,][]{Furtak2024, Wang2024, deGraaff_cliff2025}.

We adopt the model from \citet{Liu2025_sphere} to fit the NIRSpec prism spectra to describe the overall UV-to-optical continuum shape. Based on radiative-transfer calculations for a high-density spherical super-Eddington accretion flow, \citet{Liu2025_sphere} provide a library of continuum shapes that can reproduce the optical-to-NIR spectra of LRDs. We include an additional power-law component to fit the UV continuum blueward of the Balmer limit. Combining this power-law component with the spherical accretion-flow model allows us to describe the UV and optical continua, as well as the strong Balmer breaks observed in some LRDs.

As discussed in Section~\ref{results}, although the majority of sources have narrow-line ratios consistent with Case~B recombination, several sources show elevated narrow-line Balmer decrements suggestive of moderate levels of dust attenuation. Without further information regarding the geometry and distribution of dust, we assume it to be a uniform dust screen that simultaneously attenuates the UV-optical continuum, and the broad and narrow Balmer lines. Therefore, the UV and optical continua models are attenuated before comparison with the prism spectra, using the relatively flat attenuation curve of \citet{Calzetti2000} that can fit the power-law continuum observed in the rest-frame UV spectra of LRDs. Steeper attenuation curves, such as that of the Small Magellanic Cloud \citep{Gordon2003}, tend to produce a sharp drop in the bluer end of the UV continuum, which fails to match the observed spectral shape. We set $A_V$ as a free parameter that can vary within the upper and lower boundaries from the value (and uncertainties) predicted from the narrow-line Balmer decrement. 

Our final continuum model includes five free parameters: the normalization and slope of the UV power-law component, the density and luminosity of the spherical accretion flow, and the attenuation $A_V$. During the fitting, the overall model is convolved with the prism LSF corresponding to rest-frame $4000\ {\rm \AA}$. The resolving power of the NIRSpec prism varies from $R\approx 30$ to $350$ over the full wavelength coverage of $0.5-5.5\ {\rm \mu m}$. We choose the LSF at rest-frame $4000\ {\rm \AA}$ because the sharp Balmer break is more sensitive to spectral resolution than the smoother UV power-law and optical continuum.

We fit the prism spectra using MCMC within the {\tt\string GalSpec} framework and find good convergence with 10,000 sampling steps. Our goal is to obtain a practical parameterization of the UV and optical continuum shapes and to calculate the total observed luminosity of each component. We therefore integrate the best-fit power-law and optical-continuum models over $1000$--$8000\ {\rm \AA}$ to obtain the observed UV and optical luminosities. A detailed physical interpretation of the parameters in the \citet{Liu2025_sphere} model is beyond the scope of this paper. The final continuum fitting results for SID-42232, SID-68797, and SID-49140 are demonstrated in Figure~\ref{continuum_spec_fitting}.

\begin{deluxetable*}{lrrcccccccccccccccccc}[]
\label{results_table}
\tabletypesize{\footnotesize}
\tablecaption{Emission-line Measurements}
\tablewidth{0pt}
\tablehead{ID & R.A. & Decl. & $z$ & \multicolumn{2}{c}{${\rm H\alpha/H\beta}$} & \multicolumn{2}{c}{${\rm H\gamma/H\alpha}$} & \multicolumn{2}{c}{${\rm log}\ ({L_{\rm inci}}\ {\rm erg\ s^{-1}})$} & \multicolumn{2}{c}{${\rm log}\ ({L_{\rm obs}}\ {\rm erg\ s^{-1}})$} \\
\cmidrule(lr){5-6} \cmidrule(lr){7-8} \cmidrule(lr){9-10} \cmidrule(lr){11-12}
& (deg) & (deg) & & narrow & broad & narrow & broad & narrow & broad & UV & optical \\ 
(1) & (2) & (3) & (4) & (5) & (6) & (7) & (8) & (9) & (10) & (11) & (12)}

\setlength{\tabcolsep}{5pt}
\startdata
13329 & $53.1390$ & $-27.7844$ & 3.94 & $3.81_{-0.40}^{+0.49}$ & $>3.48$ & $<0.17$ & $<0.29$ & $42.53_{-0.12}^{+0.10}$ & $43.18_{-0.44}^{+0.28}$ & $44.60_{-0.21}^{+0.17}$ & $43.52_{-0.09}^{+0.08}$ \\
13704 & $53.1265$ & $-27.8181$ & 5.92 & $2.17_{-0.16}^{+0.18}$ & $>7.44$ & $0.18_{-0.04}^{+0.04}$ & $<0.17$ & $42.00_{-0.04}^{+0.03}$ & $43.09_{-0.41}^{+0.29}$ & $43.65_{-0.02}^{+0.02}$ & $43.09_{-0.03}^{+0.03}$ \\
28074 & $189.0646$ & $62.2738$ & 2.26 & $4.37$ & $13.61_{-0.49}^{+0.57}$ & \nodata & \nodata & $43.45_{-0.02}^{+0.02}$ & $43.62_{-0.06}^{+0.06}$ & $45.32_{-0.03}^{+0.03}$ & $44.56_{-0.11}^{+0.09}$ \\
38147 & $189.2707$ & $62.1484$ & 5.87 & $2.47_{-0.53}^{+0.70}$ & $16.72_{-4.14}^{+7.37}$ & $0.19_{-0.05}^{+0.06}$ & $<0.07$ & $42.59_{-0.10}^{+0.08}$ & $43.47_{-0.29}^{+0.23}$ & $44.02_{-0.03}^{+0.03}$ & $43.35_{-0.06}^{+0.05}$ \\
39353 & $189.2940$ & $62.1531$ & 4.85 & $2.85_{-0.41}^{+0.57}$ & $>2.77$ & $0.15_{-0.05}^{+0.05}$ & $<0.32$ & $41.98_{-0.04}^{+0.04}$ & $42.75_{-0.41}^{+0.27}$ & $43.67_{-0.09}^{+0.08}$ & $43.19_{-0.03}^{+0.03}$ \\
42046\textsuperscript{1} & $214.7954$ & $52.7888$ & 5.28 & $4.37$ & $11.76_{-0.99}^{+1.17}$ & \nodata & \nodata & $43.28_{-0.01}^{+0.01}$ & $44.11_{-0.31}^{+0.22}$ & $44.89_{-0.05}^{+0.05}$ & $44.44_{-0.02}^{+0.02}$ \\
42232 & $214.8868$ & $52.8554$ & 4.95 & $>1.20$ & $14.29_{-3.43}^{+6.52}$ & $<0.23$ & $<0.07$ & $41.42_{-0.13}^{+0.10}$ & $42.82_{-0.28}^{+0.22}$ & $43.37_{-0.07}^{+0.06}$ & $43.63_{-0.01}^{+0.01}$ \\
47509 & $34.2646$ & $-5.2326$ & 5.67 & $>2.10$ & $>4.13$ & $<0.26$ & $<0.14$ & $42.51_{-0.09}^{+0.08}$ & $43.35_{-0.39}^{+0.27}$ & $43.95_{-0.05}^{+0.05}$ & $43.58_{-0.05}^{+0.04}$ \\
49140\textsuperscript{2} & $214.8922$ & $52.8774$ & 6.69 & $>2.51$ & "$7.59_{-0.44}^{+0.48}$ & $<0.26$ & $0.03_{-0.01}^{+0.01}$ & $42.69_{-0.12}^{+0.10}$ & $44.23_{-0.33}^{+0.24}$ & $44.00_{-0.02}^{+0.02}$ & $44.29_{-0.01}^{+0.01}$ \\
50716 & $34.3132$ & $-5.2268$ & 6.17 & $>1.33$ & $>2.39$ & $<0.91$ & $<0.53$ & $41.44_{-0.21}^{+0.15}$ & $43.26_{-0.45}^{+0.29}$ & $43.52_{-0.22}^{+0.17}$ & $43.52_{-0.11}^{+0.09}$ \\
53501 & $189.2951$ & $62.1936$ & 3.43 & $5.02_{-0.35}^{+0.40}$ & $14.32_{-2.78}^{+4.61}$ & $0.10_{-0.01}^{+0.01}$ & $<0.07$ & $43.16_{-0.02}^{+0.02}$ & $43.36_{-0.30}^{+0.23}$ & $45.18_{-0.08}^{+0.07}$ & $43.82_{-0.03}^{+0.03}$ \\
55604 & $214.9830$ & $52.9560$ & 6.98 & $4.95_{-1.66}^{+2.43}$ & $14.17_{-0.92}^{+1.05}$ & \nodata & \nodata & $42.90_{-0.14}^{+0.10}$ & $44.02_{-0.18}^{+0.14}$ & $45.13_{-0.22}^{+0.16}$ & $44.70_{-0.09}^{+0.08}$ \\
68797\textsuperscript{1} & $189.2291$ & $62.1462$ & 5.04 & $4.37$ & $21.63_{-1.67}^{+1.91}$ & $0.09$ & $<0.02$ & $43.44_{-0.01}^{+0.01}$ & $44.01_{-0.11}^{+0.11}$ & $44.91_{-0.04}^{+0.04}$ & $44.88_{-0.01}^{+0.01}$ \\
73488 & $189.1974$ & $62.1772$ & 4.13 & $2.91_{-0.14}^{+0.15}$ & $26.49_{-5.50}^{+9.27}$ & $0.16_{-0.01}^{+0.01}$ & $<0.06$ & $42.26_{-0.02}^{+0.02}$ & $42.90_{-0.28}^{+0.22}$ & $43.77_{-0.04}^{+0.04}$ & $43.33_{-0.01}^{+0.01}$ \\
\enddata
 
\tablenotetext{1}{The narrow-line ratios are fixed to ${\rm H\alpha}/{\rm H\beta}=4.37$ and ${\rm H\gamma}/{\rm H\alpha}=0.09$.}
\tablenotetext{2}{The optical depth ratios deviate from theoretical values.}
\tablecomments{
Col. (1): Source identification from DJA.
Col. (2): Right ascension (J2000).
Col. (3): Declination (J2000).
Col. (4): Redshift.
Cols. (5)--(8): Line intensity ratios of the the narrow and broad components.
Cols. (9)--(10): Incident ionizing luminosity inferred from the narrow and broad components of \ha.
Cols. (11)--(12): Observed UV and optical continuum luminosity.  }
\end{deluxetable*}

\begin{deluxetable*}{lrrcccccccccccccccccc}[]
\label{results_table}
\tabletypesize{\footnotecccccccc}[]
\label{results_table2}
\tabletypesize{\footnotesize}
\tablecaption{Derived Physical Parameters}
\tablewidth{0pt}
\tablehead{ID & log $n_{\rm H}$ &  log $N_{\rm H}$ & $A_V$ & Break & log ($\tau_{0,{\rm H\alpha}}$) & $C_f$ & ${\rm d}v$ & $\sigma$ & log $M_{\rm BH}$ \\
& $({\rm cm^{-3}})$ & $({\rm cm^{-2}})$ & (mag) & Strength & & & $({\rm km\ s^{-1}})$ & $({\rm km\ s^{-1}})$ & ($M_{\odot}$) \\
(1) & (2) & (3) & (4) & (5) & (6) & (7) & (8) & (9) & (10)}

\setlength{\tabcolsep}{5pt}
\startdata
13329 & $9.60_{-1.32}^{+1.35}$ & $22.53_{-0.31}^{+0.37}$ & $0.99_{-0.39}^{+0.41}$ & $1.14\pm0.23$ & \nodata & \nodata & \nodata & \nodata & $6.63_{-0.08}^{+0.07}$ \\
13704 & $8.50_{-0.37}^{+0.75}$ & $22.51_{-0.32}^{+0.63}$ & \nodata & $1.12\pm0.21$ & \nodata & \nodata & \nodata & \nodata & $6.03_{-0.08}^{+0.07}$ \\
28074 & $9.59_{-0.12}^{+0.14}$ & $23.06_{-0.09}^{+0.07}$ & $1.46$ & $2.26\pm0.25$ & $>0.32$ & $0.94_{-0.01}^{+0.01}$ & $-321.30_{-4.05}^{+3.99}$ & $96.13_{-4.21}^{+6.23}$ & $7.65_{-0.09}^{+0.07}$ \\
38147 & $8.86_{-0.33}^{+0.37}$ & $22.42_{-0.26}^{+0.28}$ & \nodata & $1.15\pm0.25$ & $>0.24$ & $>0.58$ & $-279.38_{-24.91}^{+26.24}$ & $45.84_{-11.22}^{+21.77}$ & $6.51_{-0.08}^{+0.07}$ \\
39353 & $9.56_{-1.30}^{+1.34}$ & $22.53_{-0.32}^{+0.39}$ & \nodata & $1.10\pm0.27$ & \nodata & \nodata & \nodata & \nodata & $6.77_{-0.08}^{+0.08}$ \\
42046 & $10.29_{-0.52}^{+0.72}$ & $23.57_{-0.19}^{+0.13}$ & $1.46$ & $3.55\pm0.46$ & $>0.35$ & $0.74_{-0.02}^{+0.02}$ & $-500.82_{-11.54}^{+12.52}$ & $101.87_{-8.00}^{+11.15}$ & $7.65_{-0.09}^{+0.07}$ \\
42232 & $8.98_{-0.28}^{+0.79}$ & $23.40_{-0.19}^{+0.30}$ & \nodata & $3.29\pm0.46$ & \nodata & \nodata & \nodata & \nodata & $6.98_{-0.08}^{+0.07}$ \\
47509 & $9.18_{-0.95}^{+1.34}$ & $22.58_{-0.33}^{+0.39}$ & \nodata & $1.16\pm0.24$ & \nodata & \nodata & \nodata & \nodata & $6.35_{-0.09}^{+0.07}$ \\
49140 & $10.62_{-0.53}^{+0.47}$ & $23.82_{-0.15}^{+0.10}$ & \nodata & $4.92\pm0.56$ & $>0.12$ & $>0.52$ & $-22.23_{-14.69}^{+13.50}$ & $85.89_{-16.03}^{+13.58}$ & $8.03_{-0.09}^{+0.07}$ \\
50716 & $9.73_{-1.39}^{+1.31}$ & $22.61_{-0.38}^{+0.43}$ & \nodata & $1.12\pm0.51$ & \nodata & \nodata & \nodata & \nodata & $7.32_{-0.09}^{+0.08}$ \\
53501 & $10.03_{-0.54}^{+0.77}$ & $22.90_{-0.37}^{+0.28}$ & $1.94_{-0.25}^{+0.26}$ & $1.60\pm0.53$ & \nodata & \nodata & \nodata & \nodata & $7.24_{-0.08}^{+0.07}$ \\
55604 & $9.81_{-0.21}^{+0.40}$ & $23.39_{-0.13}^{+0.16}$ & $1.89_{-1.41}^{+1.38}$ & $3.36\pm0.46$ & $>0.52$ & $0.85_{-0.03}^{+0.04}$ & $-268.36_{-34.94}^{+45.21}$ & $70.09_{-10.59}^{+11.31}$ & $7.57_{-0.09}^{+0.07}$ \\
68797 & $8.86_{-0.15}^{+0.20}$ & $23.19_{-0.19}^{+0.29}$ & $1.46$ & $2.62\pm0.46$ & $>0.13$ & $0.64_{-0.02}^{+0.04}$ & $-368.72_{-7.33}^{+8.91}$ & $61.74_{-7.77}^{+13.97}$ & $7.81_{-0.09}^{+0.08}$ \\
73488 & $8.47_{-0.14}^{+0.27}$ & $22.94_{-0.44}^{+0.46}$ & $0.05_{-0.16}^{+0.17}$ & $1.40\pm0.18$ & \nodata & \nodata & \nodata & \nodata & $6.98_{-0.08}^{+0.07}$ \\
\enddata
\tablecomments{
Col. (1): Source identification from DJA. 
Cols. (2)--(3) Volume density and column density measured from the Balmer break and Balmer decrement. 
Col. (4): Dust attenuation inferred from the narrow-line Balmer decrement. 
Col. (5): Balmer break strength. 
Cols. (6)--(7): Line center optical depth of H$\alpha$ and the covering factor for the absorption component. 
Cols. (8)--(9): Central velocity offset of the absorption line relative to the narrow emission line and the intrinsic velocity dispersion. 
Col. (10): BH mass estimated using the single-epoch method of \citet{Greene_Ho_2005}.  }
\end{deluxetable*}

\section{Results}\label{results}

\subsection{Balmer Line Ratios}\label{Balmer_line_ratio}

All sources in our sample exhibit broad \ha\ emission; a broad component to \hb\ can be detected in nine sources, while \hg, on account of the low SNR, can be measured in only a single object. In terms of the narrow-line region, \hb\ can be seen in 10 sources and \hg\ in six. Figure~\ref{line_ratio_bb} presents a graphical summary of the consequent Balmer decrements, with details given in Table~\ref{results_table}.

At face value, four or five sources have broad-line Balmer decrements that formally lie close to the Case~B value (Figure~\ref{line_ratio_bb}a). Note, however, that all of values are lower limits for ${\rm H\alpha}/{\rm H\beta}$ and upper limits for ${\rm H\gamma}/{\rm H\alpha}$, such that the data points reflect more the noise properties instead of the actual line ratios. By contrast, sources with robust broad \hb\ detection have ${\rm H\alpha}/{\rm H\beta} > 7$, much higher than the decrement measured for the corresponding narrow-line component in the {\it same}\ object. In the single source for which broad \hg\ emission was successfully measured, its broad-line ratios are consistent with a significannt dust extinction of $A_{V}\gtrsim 3$ mag. 

Most sources exhibit narrow Balmer line ratios consistent with Case~B recombination modified by moderate levels of extinction ($A_{V}\lesssim 2.5$ mag). For all objects whose narrow-line Balmer decrement (lower limit) is larger than the Case~B prediction of ${\rm H\alpha}/{\rm H\beta} = 2.86$, we attribute the difference solely to dust and derive the corresponding $A_V$ and uncertainties using the \citet{Calzetti2000} attenuation curve (Table~\ref{results_table2}).

We also notice that despite the large uncertainties, the Balmer decrements of seven sources deviate from the Case~B value, but in a direction opposite to that expected from dust attenuation, pointing toward possible non-Case~B nature of the narrow lines. Overall, although the narrow-line ratios do suggest a certain amount of dust attenuation, the broad-line Balmer decrements are systematically higher than the narrow-line gas. This constitutes strong evidence that the elevated broad-line Balmer decrement in LRDs primarily does not arise from dust extinction, reaffirming the notion that dust, while perhaps not entirely negligible, does not play an energetically dominant role in defining the main physical properties of LRDs (e.g., \citealt{Chen2025_lrd_dust,Inayoshi_ho_lrd_rev}).

Intrinsic deviation from Case~B recombination is known to occur in high-density conditions that are typical to the BLR. As mentioned above, even the narrow-line gas exhibits evidence for anomalous values of the Balmer decrement that are not due to dust attenuation. In order to understand the intrinsic Balmer line ratios in regimes of moderate to high density, we adopt the results from the photoionization calculations of \citet{Yan2025}, which assume a uniform-density slab of gas with column density $N_{\rm H}=10^{22}-10^{24}\ {\rm cm^{-2}}$, metallicity $0.1\, Z_{\odot}$, and density $n_{\rm H} = 10^2-10^{12}\ {\rm cm^{-3}}$. The incident radiation continuum is parameterized as 

\begin{equation}
	f_{\nu}\propto \nu^{\alpha_{\rm UV}} e^{-h\nu/k_{\rm B}T_{\rm cut}}
	\label{def_agn_sed}
\end{equation}

\noindent
for the UV continuum and as $f_{\rm X}\propto \nu^{\alpha_{\rm X}}$ for the X-ray continuum, where $\alpha_{\rm UV}$ and $\alpha_{\rm X}$ are, respectively, the UV and X-ray spectral indices, and $T_{\rm cut}$ is the cut-off temperature of the UV bump, with $h$ Planck's constant and $k_{\rm B}$ the Boltzmann constant. \citet{Yan2025} use $\alpha_{\rm UV}=-0.5$, $\alpha_{\rm X}=-1.5$, $T_{\rm cut}=10^5\ {\rm K}$, and a spectral slope between $2500\, {\rm \AA}$ and 2 keV $\alpha_{\rm OX} = -1.5$. They also adopt a constant ionization parameter ${\rm log}\ U=-1.5$, defined as 

\begin{equation}
	U=\frac{\Phi}{n_{\rm H}c},
	\label{def_U}
\end{equation}

\noindent
where $\Phi$ is the incident ionizing photon flux, $n_{\rm H}$ the density, and $c$ the speed of light.

The broad-line Balmer ratios measured in our sample are consistent with the model predictions of \citet{Yan2025} for $n_{\rm H}\gtrsim 10^9\ {\rm cm^{-3}}$ (Figure~\ref{line_ratio_bb}a). In this density regime, the effect of dust attenuation and non-Case~B recombination are similar, as increasing $A_V$ and decreasing the density both result in larger ${\rm H\alpha}/{\rm H\beta}$ and smaller ${\rm H\gamma}/{\rm H\alpha}$. Models with density $n_{\rm H}=10^3-10^4\ {\rm cm^{-3}}$ successfully reproduce the observed values for the narrow-line Balmer decrements smaller than the values for Case~B, which otherwise cannot be explained by dust attenuation (Figure~\ref{line_ratio_bb}b). While the narrow-line Balmer decrements that exceed the Case~B value can be produced in principle by invoking high densities (blue solid line), this seems untenable because the required densities ($n_{\rm H} \gtrsim 10^{11}\ \mathrm{cm^{-3}}$) far exceed the critical density for collisional deexcitation for \OIII\ ($\sim 7\times 10^5\,\mathrm{cm}^{-3}$; \citealt{Osterbrock2006}). This underlies our choice to attribute the elevated narrow-line Balmer decrement solely to dust extinction. Although we do not find a clear correlation between the Balmer break strength and the narrow line ratios, we notice that sources with large Balmer break strength ($\gtrsim 2$) tend to be located near the position of $n_{\rm H}=10^9\ {\rm cm^{-3}}$. We will further explore this phenomenon in the next section.

\subsection{Balmer Decrement and Balmer Break}

\begin{figure}
	\centering
	\includegraphics[width=0.45\textwidth]{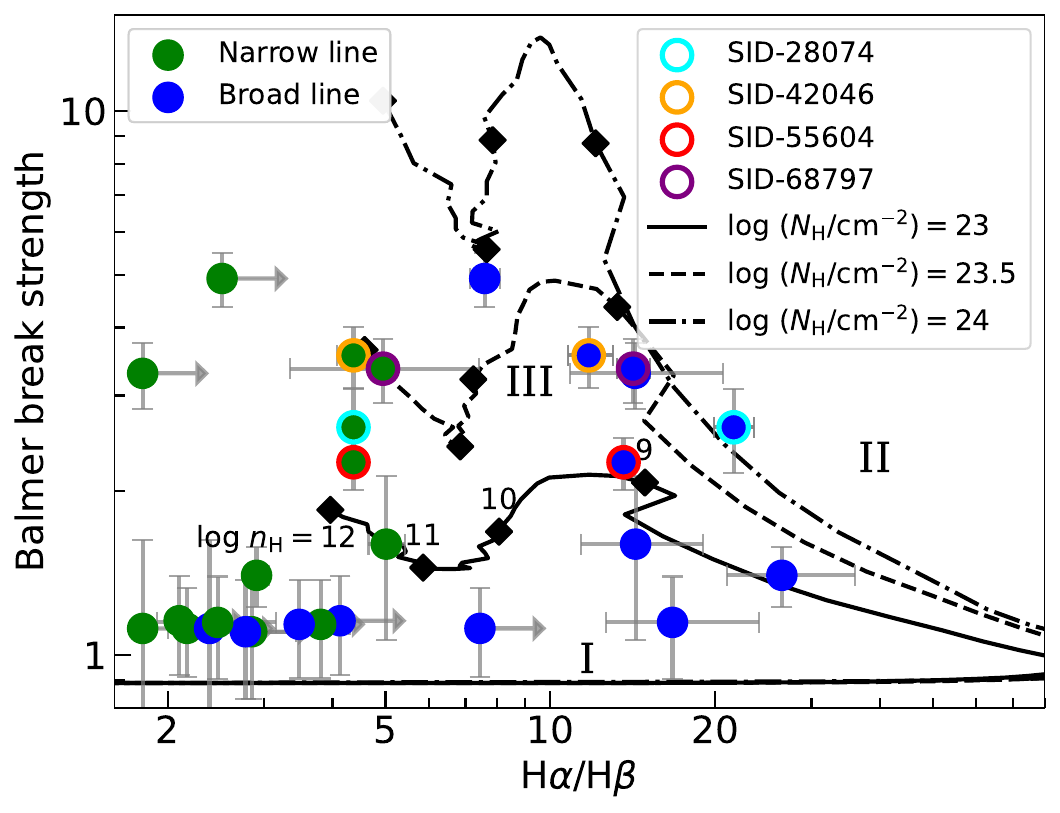}
	\caption{Variation of the Balmer decrement with the Balmer break strength. Green and blue points with errorbars represent the narrow and broad component, respectively. Points circled in cyan, orange, red, and purple mark SID-28074, SID-42046, SID-55604, and SID-68797. Theoretical curves from \citet{Yan2025} for column density $N_{\rm H}=(10^{23},\ 10^{24},\ 10^{25})\ {\rm cm^{-2}}$ are shown as black lines; black diamonds mark number density $n_{\rm H}=(10^9,\ 10^{10},\ 10^{11},\ 10^{12})\ {\rm cm^{-3}}$. }
	\label{bb_bd_cor}
\end{figure}

As predicted by \citet{Inayoshi2025_bd_ba} and \citet{Yan2025}, high-density gas can generate a Balmer break, and as density increases, the Balmer break strength correlates with the Balmer decrement (Figure~\ref{bb_bd_cor}). Both \ha\ and \hb\ become optically thick due to resonance scattering at $n_{\rm H}\gtrsim 10^5\ {\rm cm^{-3}}$. While the \ha\ photons scatter between the second and third energy level, the \hb\ photons that are scattered back to the fourth energy level can either re-emit to the second level as \hb\, or to the third level as \paa, thereby increasing the Balmer decrement. In the mean time, the density of hydrogen atoms in the second energy level is maintained by collisional excitation, which increases in proportion to $n_{\rm H}^2$. At $n_{\rm H}\lesssim 10^7\ {\rm cm^{-3}}$, these processes produce an almost horizontal line with small Balmer break (region I in Figure~\ref{bb_bd_cor}). When $n_{\rm H}\gtrsim 10^7\ {\rm cm^{-3}}$, the second energy level has accumulated enough hydrogen atoms. which start to absorb continuum photons blueward of the Balmer limit (at 3646 \AA), thus enhancing the Balmer break. Meanwhile, \paa\ becomes optically thick, and resonance scattering partially stops the channel from the fourth to the third energy level, thereby boosting \hb\ emission and reducing the Balmer decrement. The resulting curve is the increasing branch to the right side of Figure~\ref{bb_bd_cor} (region II) as the Balmer decrement decrease. As density further increases, collisional excitation moves hydrogen from the second to the third energy level and begins to suppress the Balmer break strength, generating the decreasing branch from $n_{\rm H}=10^{10}$ to $10^{11}\ {\rm cm^{-3}}$ (region III).

Our measurements show that the broad components have significantly larger Balmer decrement compare to the narrow components in the nine LRDs with both broad \ha\ and \hb\ detection. The broad-line Balmer decrement and Balmer break values can be well-reproduced by the theoretical models of \citet{Yan2025}. In Figure~\ref{line_ratio_bb} (Section~\ref{Balmer_line_ratio}, we observed that sources with large Balmer break ($\gtrsim 2$) tend to be located near a locus where $n_{\rm H} \approx 10^9\ {\rm cm^{-3}}$. This corresponds to the peak of Balmer break at density $n_{\rm H}=10^9-10^{10}\ {\rm cm^{-3}}$.

\subsection{Ionizing Budget from $H\alpha$}

LRDs emit strong Balmer emission lines. The H$\alpha$ equivalent widths in LRDs are typically three to five times larger than those observed in local AGNs \citep{Maiolino2025, Wang2025_ha_he}, implying a high luminosity of ionizing radiation. Under the standard Case~B scenario, the probability of the recombination process producing an \ha\ photon is determined by the ratio of the \ha\ recombination coefficient to the total Case~B recombination coefficient, which is 0.45 for a temperature of 10,000 K. Therefore, if the gas is optically thin to H$\alpha$ photons and is in steady state, the observed H$\alpha$ luminosity directly traces the recombination rate, which in turn equals the incident ionizing photon rate.

However, this simple Case~B conversion is not applicable to the broad H$\alpha$ emission in LRDs. As discussed in Section~4.2, LRDs typically exhibit Balmer decrements in their broad lines that are much larger than those simultaneously measured in their narrow lines, strongly suggesting that their large Balmer decrements do not arise from simple dust extinction but instead may be intrinsic to unique physical conditions. The broad-line Balmer decrements found in LRDs, H$\alpha$/H$\beta \gtrsim 8$, are also significantly higher observed in local AGNs (e.g., $\sim 3.1$; \citealt{Dong2008}). Such large Balmer decrements indicate that the broad H$\alpha$ emission is optically thick, likely due to the dense gas environment. Consequently, directly converting the observed broad H$\alpha$ luminosity into an ionizing photon rate would underestimate the intrinsic ionizing power required to produce the observed Balmer emission.

We attempt to infer the luminosity of the ionizing continuum that produces the observed H$\alpha$ emission, which hereinafter will be referred to as the incident luminosity. To account for the optically thick broad H$\alpha$ emission, we adopt the photoionization models of \citet{Yan2025}. The basic procedure is to identify, for each source, the gas column density $N_{\rm H}$ and hydrogen number density $n_{\rm H}$ that best reproduce both the observed Balmer break and the broad-line Balmer decrement. \citet{Yan2025} calculated the relation between Balmer break and Balmer decrement over the parameter ranges $N_{\rm H}=10^{22}-10^{24}\,{\rm cm^{-2}}$ and $n_{\rm H}=10^2-10^{12}\,{\rm cm^{-3}}$. We interpolate their model grid as a two-parameter function of $\log N_{\rm H}$ and $\log n_{\rm H}$ and fit the observed Balmer break and Balmer decrement using an MCMC analysis with 6000 steps. Before fitting the photoionization models, we first correct the broad-line Balmer decrements for dust attenuation, adopting the extinction inferred from the narrow-line Balmer decrement.

The Balmer decrement predicted by the models first increases and then decreases with increasing density, with the turning point located around $n_{\rm H}=10^7$--$10^8\,{\rm cm^{-3}}$. As can be seen from Figure~\ref{bb_bd_cor}, the decreasing high-density branch gradually approaches the increasing low-density branch when the column density is below $N_{\rm H} \approx 10^{23}\ {\rm cm^{-2}}$, making it hard to determine $n_{\rm H}$. We therefore constrain the density to $n_{\rm H}=10^8-10^{12}\ {\rm cm^{-3}}$ during the fitting, a range typical for BLR gas (e.g., XX give a reference), so as to only consider the high-density branch. Applying this parameter range allows us to constrain the density $n_{\rm H}$ to within 0.3 dex for sources with detectable broad \ha\ and \hb\ emission.  For the five LRDs where only lower limits on the broad-line Balmer decrement can be obtained because of the low SNR of broad H$\beta$, the uncertainty in $n_{\rm H}$ is larger, typically $\sim 0.7$ dex. These uncertainties are propagated through the following calculations using a bootstrap method.

\begin{figure}
	\centering
	\includegraphics[width=0.45\textwidth]{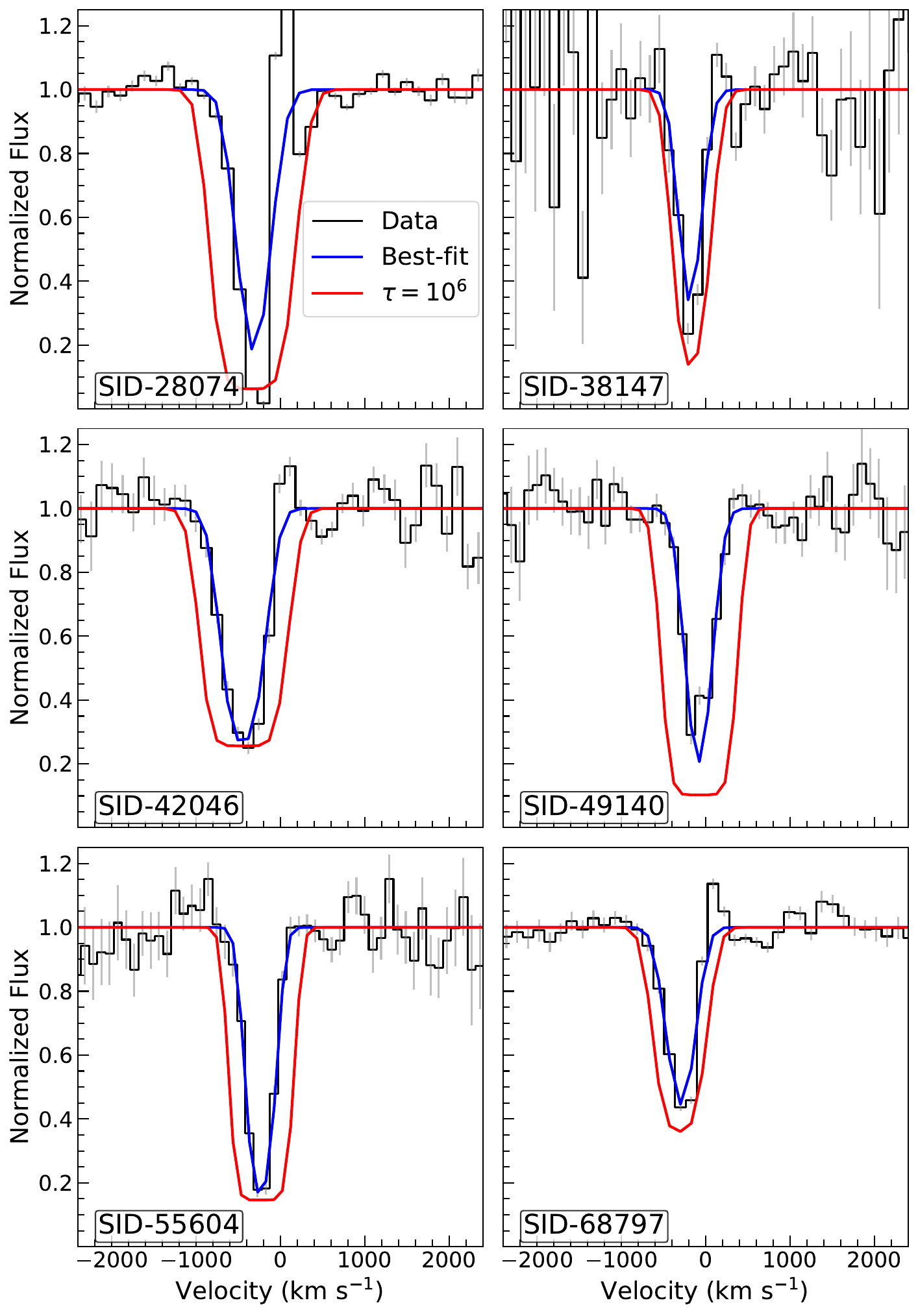}
	\caption{The absorption features in six LRDs. The observed data, plotted in black with errorbars displaying the uncertainties, are normalized by the best-fit emission-line model. The best-fit absorption-line models are shown in blue; the red lines represent the absorber with the same velocity dispersion $\sigma$ and covering factor $C_f$ but with a large optical depth of $\tau=10^6$. }
	\label{abs_trough}

\end{figure}

For a given pair of $N_{\rm H}$ and $n_{\rm H}$, \citet{Yan2025} provide the predicted H$\alpha$ flux, $f_{\rm H\alpha, m}$, for a gas slab of thickness $\Delta r = N_{\rm H}/n_{\rm H}$ and ionization parameter $\log U = -1.5$. The corresponding ionizing photon flux $\Phi$ can then be derived from the inferred gas density using Equation~\ref{def_U}.
To convert the ionizing photon flux into a total luminosity, we estimate the radial extent of the emitting gas. The outer radius is obtained by matching the observed broad H$\alpha$ luminosity $L_{\rm H\alpha, b}$ to the model flux, 

\begin{equation}
r_{\rm out} = \left( \frac{L_{\rm H\alpha, b}}{4\pi f_{\rm H\alpha, m}} \right)^{1/2}, 
\end{equation}

\noindent
while the inner radius is given by $r_{\rm in} = r_{\rm out} - \Delta r$. The total ionizing photon production rate is then $Q_{\rm ion} = 4\pi r_{\rm in}^2 \Phi$. Adopting the same SED as in \citet{Yan2025} (see Section~\ref{Balmer_line_ratio} and Equation~\ref{def_agn_sed}), we convert $Q_{\rm ion}$ into a total luminosity, defined here as the integrated emission over $1000-8000\ \mathrm{\AA}$, required to produce the observed optically thick broad H$\alpha$ emission.

\begin{figure*}
	\centering
	\includegraphics[width=1\textwidth]{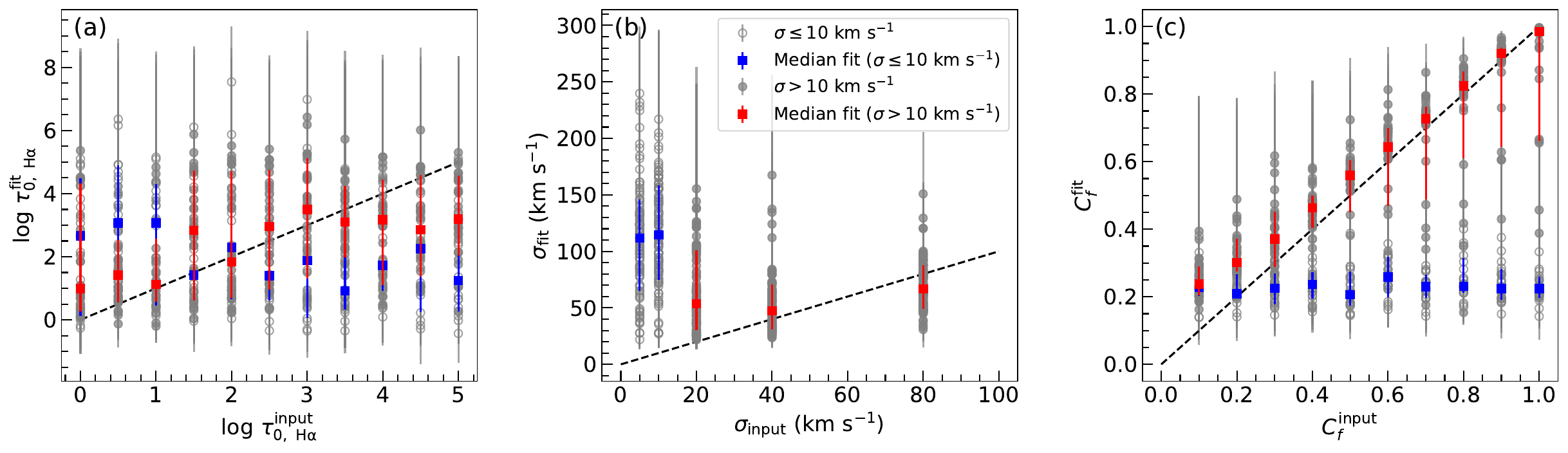}
	\caption{Comparison between the input values of (a) optical depth, (b) velocity dispersion, and (c) covering factor for the mock spectra and the best-fit absorption-line parameters for SID-42046. The open and filled grey points indicate mocks with velocity dispersion $\sigma\geq 10\ {\rm km\ s^{-1}}$ and $\sigma< 20\ {\rm km\ s^{-1}}$, respectively, with their median value and standard deviation in each bin plotted as red and blue squares.}
	\label{abs_mock_42046}

\end{figure*}

The narrow H$\alpha$ component is treated separately. In sources where the narrow-line Balmer decrement is larger than the Case~B value, we attribute the enhancement to dust extinction and correct the observed narrow H$\alpha$ flux based on the $A_V$ reported in Table~\ref{results_table2}. The total luminosity required to power the narrow H$\alpha$ emission is then estimated assuming $N_{\rm H} = 10^{22}\ \mathrm{cm^{-2}}$ and $n_{\rm H} = 100\ \mathrm{cm^{-3}}$. The implications of these results are further discussed in Section~\ref{lum_ha_obs_relation}.

\subsection{Balmer Absorption Lines}
	
\subsubsection{Absorption-line Properties}\label{tau_ratio}
	
Among the 14 LRDs with a broad-line emission, absorption features are detected in six sources (SID-28074, SID-38147, SID-42046, SID-49140, SID-55604, and SID-68797). We model the absorption line profiles with the parameterization described in Section~\ref{spec_line_model}. To break the degeneracy between the covering factor $C_f$ and line-center optical depth $\tau_0$, we simultaneously fit the line profiles for \ha, \hb, and \hg, tying the $\tau_0$ ratios according to theoretical predictions while adopting the same $C_f$ for all the lines. The optical depth ratios among different lines are consistent with the atomic physics prediction in all sources except SID-49140. For the latter case, we allowing the optical depth of \hb\ to vary freely yields a best-fit ratio of $\tau_{\rm H\alpha}/\tau_{\rm H\beta} = 0.53$, nearly 14 times smaller than the theoretical prediction. The \hg\ line can be reasonably fit by fixing the optical depth ratio to the theoretical value, but its low SNR provides only loose constraints on the absorption profile. A similar phenomenon is seen in other LRDs as well as local LRD analogs, where either the optical depth ratio disagrees with theoretical values \citep{DEugenio_2025} or the \hb\ absorption trough has a distinctive component absent from \ha\ \citep{Lin2025_local_lrd, Shangguan2026}. The physical interpretations regarding the optical depth mismatch is still unclear.	

The spectra of the six LRDs with Balmer absorption are well described by our absorption-line model after MCMC sampling with up to $10^5$ steps. However, the posterior distributions of $\tau_0$ and, in a few cases, $C_f$ remain broad, so we can only place lower limits on these parameters (Table~\ref{results_table2}). The absorbers in LRDs in general have $C_f \gtrsim 50\%$, with the one in SID-28074 reaching $C_f = 90\%$. These values are significantly larger than the values of $C_f \approx 20\%-50\%$ reported for Balmer absorption-line AGNs at $z\approx 0$ \citep{Shangguan2026}. The values of $\tau_0$ all exxceed 1, indicating optically thick absorption. Unlike the boxy absorption trough with a flat core arising from extremely high optical depth, the best-fit absorption profiles are generally centrally peaked (Figure~\ref{abs_trough}), suggesting that the optical depths are not high enough to produce a damped absorption line. 

Among the six LRDs with detected absorption lines, five (SID-28074, SID-38147, SID-42046, SID-55604, and SID-68797) show absorption profiles blueshifted by more than 200 ${\rm km\ s^{-1}}$ relative to the emission-line peaks. Interestingly, all of these sources except SID-38147 also have elevated narrow-line Balmer decrements relative to the Case~B prediction, suggesting some degree of dust attenuation\footnote{We caution that the inferred narrow-line Balmer decrements can be model-dependent, as illustrated by the alternative P-Cygni fits in Appendix~\ref{exp_pcygni}.}. In Figure~\ref{line_ratio_bb}, three of the four sources with elevated narrow-line Balmer decrements are shown as triangles on the right side of the plot because only \ha\ and \hb\ are available; the remaining source, SID-68797, is marked with a colored border. These sources are also highlighted in Figure~\ref{bb_bd_cor}. If the apparent association between strongly blueshifted absorption and dust attenuation is intrinsic, it may indicate a connection between outflowing gas and dust content on circumnuclear scales (Section~\ref{unified_model}).

\subsubsection{Robustness of the Absorption-line Fitting}\label{abs_mock}

To assess the extent to which the absorption parameters can be reliably constrained, we construct a suite of mock spectra spanning a range of absorption-line properties, following the procedures described in \citet{Shangguan2026}. For the five sources whose $\tau_0$ ratios among different transitions follow theoretical predictions, the mock spectra are generated by combining the best-fit continuum, broad-line and narrow-line emission components, and an input absorption component, with noise drawn from the pixel-wise flux uncertainties. For SID-49140, whose optical depth ratios are inconsistent with theoretical values, we slightly modify the procedures to generate mock spectra using the best-fit optical depth ratios. We note that the results for SID-49140 should be interpreted with caution because of the unclear physical reason for the optical depth ratio mismatch. 
We vary the input absorption parameters over a grid defined by optical depth $\log \tau_{\rm 0, H\alpha}$ (11 values uniformly spaced between 0 and 5), velocity dispersion $\sigma$ (5, 10, 20, 40, and 80 ${\rm km\ s^{-1}}$), and covering factor $C_f$ (10 values uniformly spaced between 0.1 and 1.0), yielding a total of 550 mock spectra. The velocity dispersion and covering factor are assumed to be identical for the H$\alpha$, H$\beta$, and H$\gamma$ absorption lines, while the optical depths are tied according to theoretical ratios. Each mock spectrum is fitted using 20,000 sampling steps; increasing the number of steps further does not lead to qualitative changes in the results. 

The comparison between input and recovered parameters for SID-42046 is shown in Figure~\ref{abs_mock_42046}, while the results for the other four sources are presented in the Appendix. We find that the recovery of absorption parameters depends sensitively on the velocity dispersion. For $\sigma \geq 20\ {\rm km\ s^{-1}}$, the velocity dispersion is generally well recovered, with only minor deviations at $\sigma = 20\ {\rm km\ s^{-1}}$. In contrast, for $\sigma < 20\ {\rm km\ s^{-1}}$, the recovered values exhibit significant systematic overestimation. A similar trend is observed for the covering factor: for $\sigma \geq 20\ {\rm km\ s^{-1}}$, $C_f$ can be robustly constrained, particularly for $C_f \gtrsim 0.5$, whereas it becomes effectively unconstrained at lower velocity dispersions mainly because the absorption lines cannot be detected when the absorption line is too narrow. The optical depth, however, is poorly recovered across the full parameter space, with large deviations from the input values observed at all $\sigma$. This results from a combination of low spectral resolution blurring the absorption profile and low SNR in the broad \hb\ and \hg\ lines lacking information to constrain the optical depth. Notably, all three sources analyzed here have best-fit values of $\sigma \geq 40\ {\rm km\ s^{-1}}$, indicating that the inferred (lower limits on) covering factors are robust. In contrast, the optical depth measurements should be treated with caution.

\section{Discussion}\label{discussion}

\begin{figure*}
	\centering
	\includegraphics[width=1\textwidth]{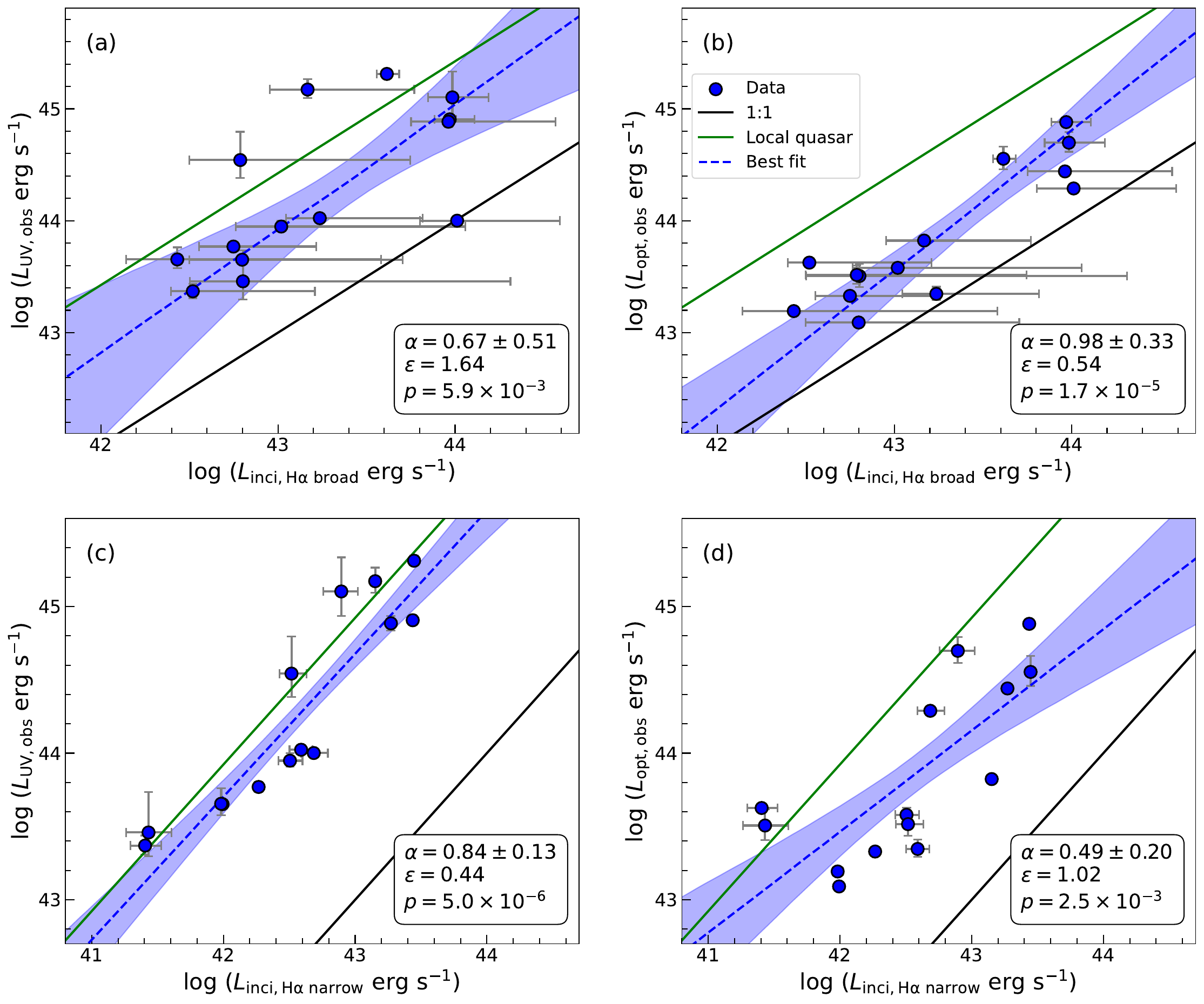}
	\caption{The observed UV and optical luminosity as a function of the incident luminosity predicted from broad and narrow \ha\ emission. Green solid line represents the relation between observed luminosity (integrated from $1000-8000\ {\rm \AA}$) and the incident luminosity predicted from the composite spectrum of local quasars \citep{VandenBerk2001}. Black solid line gives the 1:1 relation. Blue dashed line shows the best-fit relation, with the uncertainties denoted by the blue shaded region; the best-fit power-law index $\alpha$, intrinsic scatter $\epsilon$, and significance ($p$-value) of Pearson's correlation are labeled in the panels.}
	\label{uv_ha_energy}
\end{figure*}

	\subsection{Correlations Between the Observed and Incident Luminosity}\label{lum_ha_obs_relation}

The blackbody-like optical-NIR SEDs of LRDs are widely thought to arise from the reprocessing of high-energy photons by dense, optically thick gas \citep[e.g.][]{Inayoshi_ho_lrd_rev}. The dense medium populates the $n=2$ level of hydrogen through collisional excitation and Ly$\alpha$ photon trapping, thus producing relevant continuum and line absorption features \citep{Juodzbalis2024, Inayoshi2025_bd_ba}. At sufficiently high optical depths, the emergent emission may approach a blackbody-like spectrum, potentially explaining the observed optical--NIR SED \citep{Kido2025, Liu2025_sphere, Inayoshi_Shangguan2026, Naidu2025_bh_star}. However, the relative geometry among the ionizing continuum source, the H$\alpha$-emitting clouds, and the dense gas responsible for the optical continuum remains highly uncertain.

On the one hand, the line-emitting clouds must be sufficiently close to the ionizing source to produce the observed H$\alpha$ luminosity. On the other hand, the H$\alpha$ photons generated via recombination must be able to escape the surrounding dense medium to be observable. Together, these competing requirements impose stringent constraints on the geometry and physical coupling among the ionizing source, the broad-line and narrow-line region, and the medium emitting the red optical continuum. To investigate this configuration, we compare the incident luminosities inferred from the broad and narrow H$\alpha$ emission with the attenuation-corrected UV and optical continuum luminosities (Figure~\ref{uv_ha_energy}), aiming to clarify how they are related to the UV and optical part of the ``V-shaped'' SED. A Pearson correlation analysis reveals statistically significant correlations ($p < 0.05$) between the UV as well as optical luminosity and the incident luminosities inferred from both the broad and narrow H$\alpha$ emission. The incident luminosity inferred from the broad-line emission correlates more significantly with the optical luminosity than with the UV luminosity, whereas that inferred from the narrow-line emission shows a stronger correlation with the UV luminosity. These trends suggest distinct connections between the different line-emitting components and the sources of the UV and optical continuum emission.

For comparison, we also calculate the ratio between the observed luminosity and the incident luminosity for the broad and narrow \ha\ emission in the local quasar composite spectrum of \citet{VandenBerk2001}. The ratio of $1500\ {\rm \AA}$ luminosity to H$\alpha$ luminosity is 70 for the broad component and 220 for the narrow component \citep{Asada2026}. Based on these ratios, we convert \ha\ luminosity to $Q_{\rm ion}$ using the Case~B radiative coefficient at 10,000 K \citep{Draine2011}, and then estimate the continuum luminosity from $Q_{\rm ion}$ using the procedure adopted for the LRDs (Section~\ref{Balmer_line_ratio}), assuming $\alpha_{\rm UV}=-0.44$ \citep{VandenBerk2001}.

The incident luminosities inferred from H$\alpha$ emission are expected to scale linearly with the observed luminosities if both components are photoionized by accretion-disk radiation. For reference, we plot a linear relation in each panel, with normalization anchored to the ratios of local quasars (solid green lines in Figure~\ref{uv_ha_energy}). We perform linear regression analysis for the observed and incident luminosities of different components in our sample using the package {\tt\string linmix} \citep{Kelly2007}, which considers the uncertainties in both the dependent and independent variables and fits for the slope $\alpha$, zero point $b$, and intrinsic scatter $\epsilon$. The incident luminosity of LRDs inferred from broad H$\alpha$ correlates nearly linearly with both the UV ($\alpha=0.66\pm 0.53$; Figure~\ref{uv_ha_energy}a) and optical ($\alpha=0.97\pm 0.33$; Figure~\ref{uv_ha_energy}b) continuum luminosity, with a best-fit power-law index consistent with unity within $1\, \sigma$ uncertainties. The correlation is stronger with the optical luminosity ($p=1.7\times 10^{-5}$) and weaker for the UV luminosity ($p=5.9\times 10^{-3}$). The similarity with the slope found in local quasars suggests that the optical continuum and broad H$\alpha$ emission in LRDs are closely coupled through photoionization. Of course, the optical continuum itself is unlikely to be the direct source of ionization. Its blackbody-like SED with an effective temperature $\sim 5000\ {\rm K}$ produces negligible hydrogen-ionizing photons. A more plausible interpretation is that both the optical continuum and the broad H$\alpha$ emission are powered by the same underlying ionizing source. In this scenario, most of the incident ionizing radiation is absorbed by an optically thick gaseous medium and re-emitted at optical wavelengths, approximately conserving the total energy budget, while a fraction of the ionizing photons directly illuminates the BLR to power the observed broad H$\alpha$ emission. The same ionizing continuum that travels toward our line-of-sight is detected in the UV, thus accounting for the observed near-linear correlation between the UV continuum and broad H$\alpha$ emission.

The incident luminosity from narrow H$\alpha$ emission also correlates strongly with the UV luminosity (Figure~\ref{uv_ha_energy}c). Indeed, most data points lie close to the relation of local quasars, both in slope ($\alpha=0.84\pm 0.13$) and normalization,  suggesting that the production mechanism for the narrow-line emission in LRDs is similar to that in conventional AGNs. By contrast, the correlation between narrow H$\alpha$ and the optical continuum luminosity is much weaker (Figure~\ref{uv_ha_energy}d), showing a $p$-value $\sim 500$ times larger compare to the correlation with UV luminosity as well as large intrinsic scatter. The power-law slope of $\alpha=0.48\pm 0.21$ is slightly lower but still consistent within $3\, \sigma$ confidence level with unity.

\begin{figure*}
	\centering
	\includegraphics[width=1\textwidth]{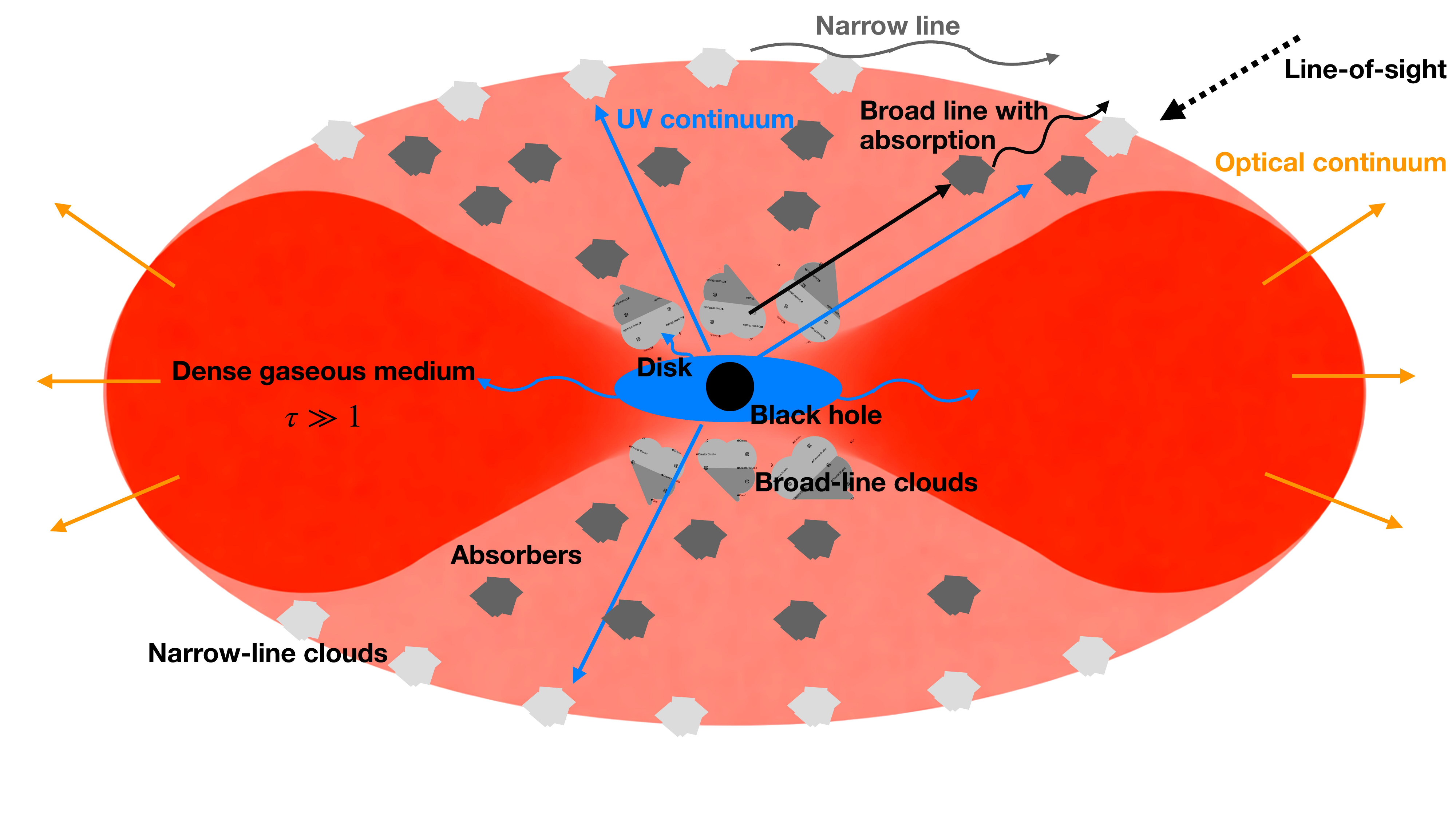}
	\caption{A conceptual illustration of ``clumpy gaseous torus'' nuclear structure of LRDs. The BH (black dot) and its accretion disk (blue oval) are enshrouded in an ambient gas reservoir (regions in red). The density of the gas reservoir is angle-dependent, being highest along the major axis of the accretion disk (dark red) and lowest along the polar direction (light red). The high-density gas constitutes the optically thick, dense, gaseous medium that emits the blackbody-like optical continuum. The UV continuum from the accretion disk escapes along the poles. High-density gas clumps (dark gray) are embedded in the low-density ambient gas; those closer to the BH produces the broad Balmer lines, while ones farther away are the absorbers that generate the Balmer absorption troughs. The light gray clouds at at larger distances produce the narrow component of Balmer emission as well as forbidden lines.}
	\label{final_toy_model}
\end{figure*}

\subsection{An Overall Picture}\label{unified_model}

Can we construct an overall picture that can self-consistently explain the key spectral features discussed above? We recap the main pieces of observational evidence: 

\begin{itemize}

\item (A) Balmer absorption is detected in six out of the 14 LRDs, all of which possess a centrally peaked, Gaussian-like absorption trough (Figure~\ref{abs_trough}), but none showing the boxy shape characteristic of damped absorption in extremely optically thick systems. 

\item (B) A Balmer break is clearly present in all sources. 

\item (C) The V-shaped continuum can be fitted with the combination of a power law in the UV and the spherical dense gas model of \citet{Liu2025_sphere} in the optical, both components subject to moderate correction for dust attenuation based on the narrow-line Balmer decrement (Figure~\ref{continuum_spec_fitting}).

\item (D) The detection and simultaneous spectral decomposition of the \ha\, \hb\ and \hg\ lines reveals broad-line Balmer decrements that are systematically larger than those of the narrow lines, favoring the dense gas processes instead of dust reddening as the principal cause for elevating the broad-line ratios (Figure~\ref{line_ratio_bb})

\end{itemize}

Combining points (B) and (D) with predictions from the photoionization calculations of \citet{Yan2025}, we infer the incident continuum luminosity required to produce the observed \ha\ emission under optically thick conditions. The significant correlation (Figure~\ref{uv_ha_energy}) between the incident luminosity of the broad and narrow \ha\ line emission with the extinction-corrected UV and optical continuum luminosity (point C)---one that mirrors a similar empirical trend seen in canonical local AGNs---suggests that photoionization is likely the primary mechanism linking the Balmer lines and the observed continuum.

To decipher points (A)--(D) under a coherent physical picture for the nuclear structure of LRDs, we start with the tight relation between the optical continuum and broad \ha\ luminosity, one that is consistent with photoionization. There are two ways to interpret this correlation. The first scenario, proposed by \citet{Pang2026}, posits that the broad lines arise from discrete clouds blended with a similarly clumpy, dense gaseous medium. Both the BLR clouds and the dense gas are illuminated by a co-spatial accretion disk emitting UV emission, leading to the correlated broad-line luminosity and the reprocessed optical continuum from the dense gaseous medium. The second scenario, conceptually illustrated in Figure~\ref{final_toy_model}, places the BLR clouds above the disk surface. The dense gaseous medium constitutes a natural extension of the accretion disk to larger radii, both being optically and geometrically thick, allowing it to absorb and reprocess the accretion disk continuum. The major difference between the two scenarios is that in the former the broad Balmer emission travels through the optically thick gaseous medium before reaching the observer because of the spatial blending, while in the latter situation the broad emission lines reach the observer through a different, relatively less obscured line-of-sight. 

The most important clue that we have at our disposal to distinguish between these two scenarios is the profile of the absorption lines. The effective optical depth for H$\alpha$ absorption by hydrogen in the $n=2$ level is $\sim 10^6$ times larger than the continuum optical depth at the Balmer edge at $3646\ \mathrm{\AA}$ \citep{Draine2011}. If the broad-line clouds were co-spatial with this medium, escaping H$\alpha$ photons would undergo severe absorption, potentially producing a strongly suppressed or boxy damped absorption profile. Although our absorption-line modeling provides only lower limits on the optical depth, comparison with a model assuming $\tau_{\rm H\alpha}=10^6$ shows a clear mismatch with the observed absorption trough (Figure~\ref{abs_trough}). This indicates that the broad-line clouds are unlikely to be embedded within the dense optically thick medium; otherwise, the true optical depth would be even larger, worsening the discrepancy. To alleviate tension with the first scenario would require fine-tuning of the covering factor for individual cloud and the clumpiness of the dense gaseous medium in order to match with the observed absorption-line profile. The evidence presented in this study strongly favors the nuclear structure illustrated in Figure~\ref{final_toy_model}, which we designate the ``clumpy gaseous torus.''

The clumpy gaseous torus framework naturally accounts for the correlation between UV and broad \ha\ luminosity because the observed UV continuum is the ionization source for the BLR. Clouds at larger radii along the polar direction that are ionized by the escaping UV radiation produce the narrow emission lines, and hence explains the correlation between UV and narrow \ha\ luminosity. An AGN origin for the UV continuum is also supported by the detection of \ion{Fe}{2} emission and high-ionization lines such as \ion{N}{5} in some LRDs \citep{Akins2025_lrd_uvlines, Tang2025, PerezGonzalez2026}. The overall weakness of the high-ionization metal lines, along with the low \ion{He}{2}$/{\rm H\alpha}$ ratio \citep{Wang2025_ha_he}, likely reflects the low-metallicity environment of LRDs \citep{Ivey2026, Korber2026, Maiolino2026_metallicity}, an intrinsically softened accretion disk continuum \citep{Kubota2018, Kubota2019, Inayoshi2024_xray}, or a certain level of host galaxy contribution to the UV continuum \citep{Chen2025, Chen2025_lrd_dust, Zhuang2025_lrd}.

If the BLR and the dense gaseous medium are spatially entangled, what produces the Balmer absorption lines? Absorption features are detected in six LRDs ($43\%$) in our sample, a higher frequency than previous estimates of $10\%-20\%$ \citep{Matthee2024_lrd_wfss, Lin2025_hae, Zhuang2025_lrd} presumably because we intentionally included four additional sources with prominent absorption features that only have \ha\ and \hb\ coverage. From our line-profile modeling, we find that all detected absorbers have covering factors exceeding $50\%$, with a maximum value of $90\%$ in SID-28074. Unlike the marginally resolved narrow emission-line component, the mock analysis in Section~\ref{abs_mock} indicates that the absorption features are spectrally resolved with well-constrained velocity dispersions, suggesting that the absorbers occupy a region of intermediate spatial scale between the broad-line and narrow-line emission zones.
    
Several conclusions can be drawn regarding the physical properties of the absorbers based on these measurements. The fraction of sources exhibiting Balmer absorption provides a first-order constraint on the incidence of favorable absorbing sightlines, which is on average $10\%-20\%$. This quantity should be distinguished from the covering factor inferred for individual detected absorbers, which describes the fraction of the background emission covered along those particular lines of sight. The apparent tension between the modest detection rate and the relatively high covering factors inferred in individual systems likely reflects strong selection effects. Balmer absorption is only detectable under favorable conditions, where the absorber has a sufficiently large velocity dispersion and a large flux contrast at line center, making systems with higher covering factors preferentially identified. Overall, the covering factor is expected to vary across the population but remain below unity, implying a clumpy instead of a continuous absorbing medium.

We observe an interesting trend between the velocity shift of the absorption line and dust attenuation. The majority of sources with blueshifted absorption lines (four out of five) have elevated narrow-line Balmer decrement relative to Case~B, suggesting the presence of dust attenuation. Radiation pressure on dust grains is known to be an efficient driver for outflows \citep{Ishibashi2015}. The high dust opacity effectively couples the accretion disk emission and circumnuclear gas, providing stronger radiation pressure launch outflows that subsequently absorbs emission from the BLR. LRDs with blueshifted absorption lines could represent a relatively evolved phase in which metal enrichment from central star formation produces dust on circumnuclear scales, which in turn accelerates gas outflows. 

Overall, our scenario is qualitatively similar to the conceptual framework proposed by \citet{Lin2025_local_lrd}, in which a non-spherical distribution of dense gas gives rise to viewing angle-dependent observational properties across the LRD population. Most LRDs are selected based on their characteristic V-shaped SEDs, which require the simultaneous detection of both UV and optical emission \citep[e.g.,][]{Greene2024_lrd_spec, Kocevski2025_lrd}. This selection criterion effectively biases the sample toward systems viewed from relatively polar orientations, such that the ionizing UV continuum from the accretion disk remains at least partially observable. In contrast, the UV continuum would be strongly attenuated or completely obscured from more edge-on viewing angles, leaving predominantly the reprocessed optical emission. Such a configuration may apply to the source reported in \citet{Naidu2025_bh_star}, whose SED is well described by a nearly pure blackbody-like continuum with no detectable UV component. The associated H$\beta$ line profile exhibits a peculiar shape that can be modeled as a single Gaussian affected by multiple scattering, consistent with photon propagation through an optically thick medium \citep{Laor2006, Chang2025}. Similarly, \citet{Matthee2026_lrd_prof} find that H$\alpha$ line profiles vary systematically with optical-to-UV color, where redder and optically brighter sources tend to show absorption-dominated cores and more extended exponential wings, consistent with scattering effects in a dense medium. These results suggest that lines of sight with higher optical-to-UV luminosity ratios intersect increasingly more optically thick material responsible for both absorption and scattering in the Balmer lines. Taken together, these lines of evidence support a unified nuclear structure in which viewing angle simultaneously regulates the relative contribution of reprocessed and direct continuum emission, as well as the degree of absorption and scattering affecting both the continuum and Balmer line profiles.

\subsection{Stratification of the Broad-line Region}\label{stratified_gas}

\begin{figure*}
	\centering
	\includegraphics[width=1\textwidth]{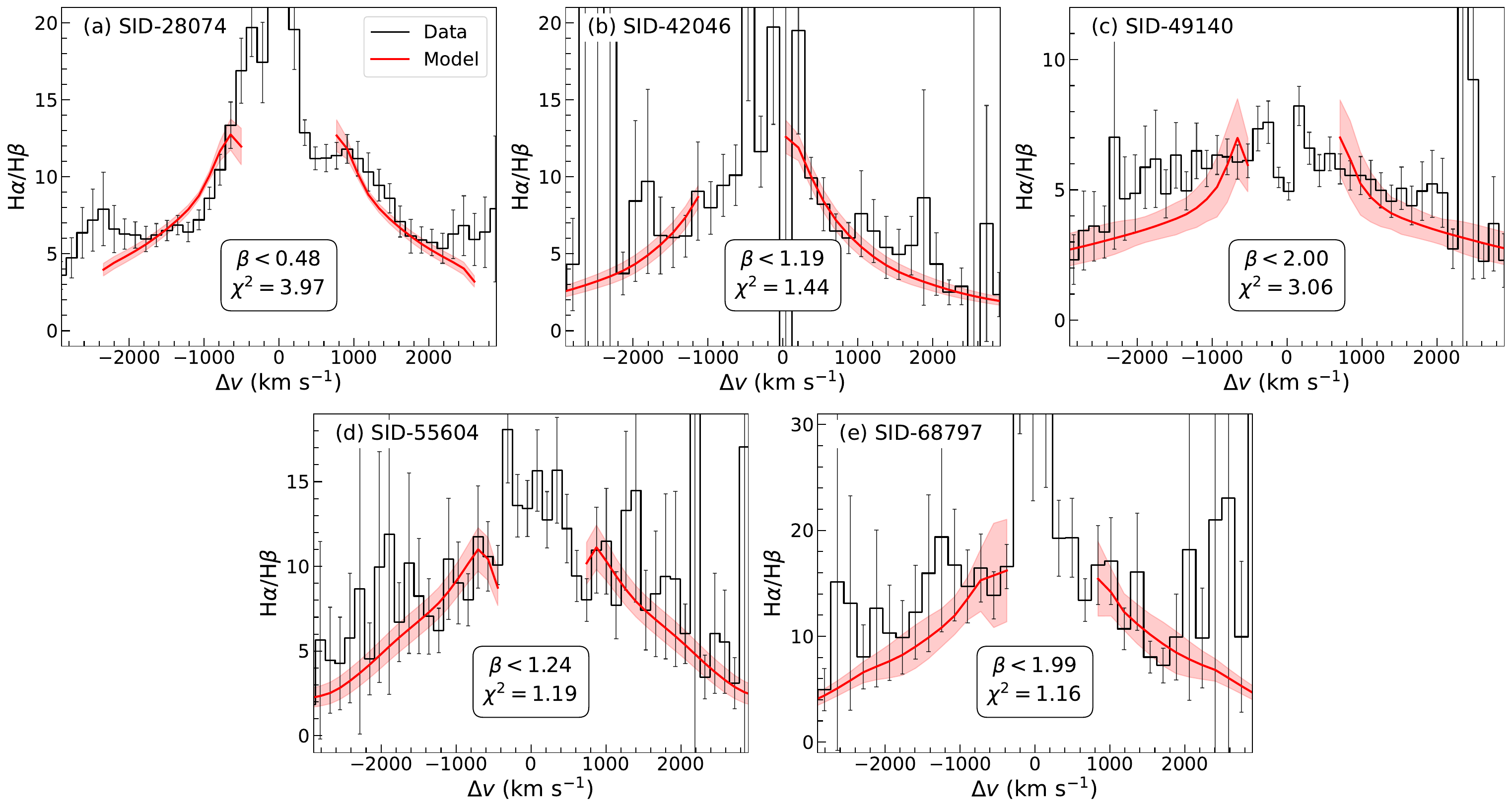}
	\caption{Balmer decrement as a function of velocity for the five sources with average $\rm SNR > 3$: (a) SID-28074, (b) SID-42046, (c) SID-49140, (d) SID-55604, and (e) SID-68797. The best-fit stratified gas model is plotted as a red solid curve, with uncertainties shown by the shaded regions. The legend gives the reduced $\chi^2$ of the fit as well as the power-law index for the density profile ($\beta$). }
	\label{bd_vel_res}
\end{figure*}

In a virialized BLR, the velocity correlates with radial distance from the BH as $v\propto r^{-0.5}$. Gas at larger distances contributes primarily to the low line-of-sight velocity core of the emission line, while the high-velocity wings originate from gas closer to the BH. Consequently, the Balmer decrement as a function of velocity provides insights into the stratification of gas properties within the BLR and potentially can constrain the gas distribution. In this section, we measure the velocity-resolved Balmer decrement profiles and then construct a simple stratified BLR model based on photoionization calculations and virialized clumpy gas to reproduce the observed velocity-dependent Balmer decrement profiles. This model allows us to investigate whether the observed Balmer decrement gradients can be explained by radial variations in gas density and ionization parameter within the BLR.

\subsubsection{Velocity-resolved Balmer decrement}

To measure the velocity-resolved Balmer decrement, we first subtract the best-fit continuum and narrow-line components from the spectra. Any flux lost to absorption, if present, is added back to the continuum- and narrow line-subtracted data to obtain the final broad line-only spectra for \ha\ and \hb. The spectra with narrower LSFs, typically \ha, are then convolved with a Gaussian kernel to match the resolution of the broader line. Finally, the convolved spectra are interpolated and reprojected onto a common velocity grid, with uncertainties propagated throughout the process.

Only five sources in our sample (SID-28074, SID-42046, SID-49140, SID-55604, and SID-68797) have sufficient SNR to yield a meaningful velocity-resolved Balmer decrement profile (Figure~\ref{bd_vel_res}). In all cases, the Balmer decrement peaks in the central, low-velocity core and decreases by at least a factor of 2 toward the high-velocity wings, resembling trends observed in some local AGNs \citep[e.g.][]{Feng2024, Li2024_bd_vel}. As demonstrated by velocity-resolved reverberation mapping studies \citep{Li2024_bd_vel}, such centrally peaked and approximately symmetric structures are indicative of a predominantly virialized BLR, in which bulk inflow or outflow motions are subdominant and velocity inversely traces distance from the BH. In this framework, the observed variation of the Balmer decrement with velocity reflects spatial changes in physical conditions, such as density and ionization parameter, across the BLR, thereby providing constraints on its radial stratification.

\subsubsection{Stratified BLR model and fitting results}

To further investigate the structure of the broad-line gas, we construct a model based on photoionization calculations aimed at reproducing the observed velocity-dependent Balmer decrement profiles. For LRDs, constraints on the BLR column density $N_{\rm H}$ are derived from the Balmer break strength (Figure~\ref{bb_bd_cor}), where larger $N_{\rm H}$ corresponds to a stronger Balmer break. In our model, the broad-line gas is represented as an ensemble of discrete clumps distributed over a range of radii from the BH. The clump density follows a radial power-law profile, $n_{\rm H} \propto r^{-\beta}$. Each clump is assigned a fixed column density $N_{\rm H} = 10^{22}\ \mathrm{cm^{-2}}$, and the number of clumps is adjusted such that the integrated line-of-sight column density matches the value inferred from the Balmer break.

We assume purely virial kinematics for the clumps, with velocities given by $v = (GM_{\rm BH}/r)^{1/2}$. The BH masses are estimated from the luminosity and FWHM of the broad H$\alpha$ line using the calibration of \citet{Greene_Ho_2005}. The clumps are distributed to uniformly sample velocity space and are truncated at inner and outer boundaries corresponding to $v = 4000\ \mathrm{km\ s^{-1}}$ and $500\ \mathrm{km\ s^{-1}}$, respectively, consistent with the observed broad wings of the H$\alpha$ profiles. Assuming photoionization by a central radiation source, the ionization parameter scales as $U \propto r^{-2}$.

Calculating the overall Balmer decrement profile requires photoionization modeling of gas over a range of densities and ionization parameters. The calculations of \citet{Yan2025} provide Balmer decrements for low-metallicity gas over a range of densities applicable to LRDs, but only at a fixed ionization parameter of $\log U=-1.5$. To account for the contribution from gas clumps at larger radii, where the incident photon flux and hence the ionization parameter are expected to be lower, we incorporate another set of calculations from \citet{Korista1997}, who adopt a similar setup as \citet{Yan2025} (see Section~\ref{Balmer_line_ratio}). The main differences are that \citet{Korista1997} assume solar metallicity, which is more applicable to low-redshift AGNs, and they compute models over a grid of incident photon fluxes, $\Phi = 10^{17}$--$10^{24}\ {\rm cm^{-2}\ s^{-1}}$, which correspond to different ionization parameters.

For a fixed gas density, the Balmer decrement increases as the ionization parameter decreases. This trend is mainly driven by two effects. First, a lower ionization parameter changes the electron temperature, which affects the emissivities of the \ha\ and \hb\ transitions. Second, it increases the neutral fraction of hydrogen, making both \ha\ and \hb\ more optically thick through enhanced Ly$\alpha$ photon trapping. To combine the low-metallicity results of \citet{Yan2025} with the ionization-parameter dependence from \citet{Korista1997}, we make the simplifying assumption that the relative dependence of the Balmer decrement on ionization parameter is similar at different metallicities. We therefore adopt the trend between the Balmer decrement and ionization parameter from \citet{Korista1997}, but renormalize the absolute Balmer decrement so that the value at $\log U=-1.5$ matches that from \citet{Yan2025} at $Z = 0.1\,Z_{\odot}$.

Our model predicts velocity-resolved Balmer decrement profiles using four free parameters: the density power-law slope $\beta$, the ionizing photon flux normalization $\log \Phi_{\rm in}$, the density normalization $\log n_0$ at the inner boundary, and a velocity shift parameter $v_{\rm shift}$ to account for small systematic offsets in the profile centroid. We convolve the resulting model profiles with the instrumental resolution, and we use MCMC sampling to fit the observed Balmer decrement profiles of the five high-SNR LRDs (Figure~\ref{bd_vel_res}).

The model provides a satisfactory description of the observed Balmer decrement profiles for SID-42046, SID-55604, and SID-68797, resulting in reduced $\chi^2$ values of 1.44, 1.19 and 1.16, respectively. The inferred density slope $\beta$ is only weakly constrained, giving only a $3\, \sigma$ upper limit. The density slope is $\beta < 1.19$ for SID-42046 and $\beta < 1.24$ for SID-55604, systematically lower than $\beta = 1.5$ expected for an idealized spherical \citet{Bondi1952} accretion flow and also lower than those inferred for local Seyfert 1 galaxies, typically $1 \lesssim \beta \lesssim 1.5$ \citep{Kaspi1999}. The upper limit for SID-68797, $\beta<1.99$, is larger. In contrast, the model does not reproduce well the Balmer decrement profiles of SID-28074 and SID-49140, with reduced $\chi^2$ values of 3.86 and 3.06, respectively. The more irregular profile of SID-28074 suggests departures from a simple virial interpretation. This source also has the largest narrow-line Balmer decrement in our sample, ${\rm H}\alpha/{\rm H}\beta\approx 6$, corresponding to $A_{V}=2.56$ mag. Its absorption line is blueshifted and has the highest covering factor among the absorption-line sources, $C_f=0.90$. The mismatch with the virialized model may therefore reflect the influence of strong outflows, which could perturb the BLR dynamics and connect to the larger-scale dust-rich outflow traced by the blueshifted absorption. By contrast, SID-49140 shows a much shallower decline from the line core to the wings. It also exhibits the strongest Balmer break and the deepest absorption trough in our sample. Within the framework discussed in Section~\ref{unified_model}, this discrepancy may indicate that the observed line profile of SID-49140 is significantly affected by scattering processes \citep[see also][]{Nikopoulos2025}, and therefore may not directly trace the intrinsic BLR kinematics.

\section{Summary}

LRDs are known to exhibit distinctive SEDs compared to AGNs in the local Universe, suggesting differences in their nuclear structures that may deviate from the conventional paradigm. The profiles and intensities of the Balmer lines potentially reflect the physical conditions of the nuclear gas. We comprehensively analyze the low-resolution and medium-resolution JWST/NIRSpec spectra of 14 LRDs to investigate their Balmer absorption lines, Balmer break, UV and optical continua, and Balmer decrement (hence ABCD). By correlating the emission-line luminosities with the continuum luminosities, we derive a model for the nuclear structure that can reproduce the measurements of both the emission and absorption features. With the aid of photoionization calculations, we investigate both the integrated and velocity-resolved line ratios with the goal of constraining the gas properties. Our main conclusions are summarized as follows.

\begin{itemize}

\item The broad components of the emission lines generally show much larger Balmer decrements than the narrow components. When \ha/\hb\ and \hg/\ha\ are considered simultaneously, the line ratios of the narrow components are consistent with low to moderate levels of dust extinction. The presence of a significant Balmer break, in conjunction with the modest attenuation inferred from the narrow lines, strongly suggests that the elevated Balmer decrements observed in the broad lines arise not from dust extinction but instead reflect physical processes associated with high-density gas.

\item Balmer absorption features are detected in six LRDs. The absorption troughs generally show centrally peaked profiles and do not exhibit the boxy shapes commonly seen in damped absorption systems. Simultaneously fitting the absorption profiles of \ha, \hb, and \hg, we find that the absorption troughs in five LRDs can be well reproduced by tying the kinematics and covering factor of the absorber and fixing the optical depth ratios to their theoretical values. The exception is SID-49140, for which the optical depth ratio between \ha\ and \hb\ deviates significantly from the theoretical prediction.

\item Sources with blueshifted absorption lines tend to have elevated narrow-line Balmer decrements. This trend, if not an artifact of the model-dependence of the narrow-line decomposition, may potentially indicate a connection between Balmer absorption and dust-driven outflows in LRDs.

\item Similar to conventional AGNs, the incident luminosity inferred from the extinction-corrected broad and narrow \ha\ emission strongly correlates with both the UV and optical continuum luminosities, suggesting that the observed continuum and Balmer emission lines are linked through photoionization.

\item We propose a viewing-angle-dependent nuclear structure in which an optically thick, clumpy gaseous torus surrounds the central accretion disk, while the line-emitting gas is directly ionized along relatively unobscured polar directions. 

\item Analysis of the velocity-resolved Balmer decrements for sources with sufficiently high-quality spectra allow us to constrain the power-law index of the radial density profile of the BLR to $\beta < 2$.

\end{itemize}

\begin{acknowledgments}
This work was supported by National Key R\&D Program of China (2022YFF0503401), the China Manned Space Program (CMS-CSST-2025-A09), the National Science Foundation of China (12233001, 12573014), the National SKA Program of China (2025SKA0130100), and the Fundamental Research Funds for the Central Universities, Peking University (7100604896). CHC thanks Zu Yan, Kirk Korista, and Yoshihisa Asada for helpful comments and discussions.
\end{acknowledgments}

\appendix

\section{Spectral Fitting Results}

Simultaneous \ha, \hb, and, if present, \hg\ spectral fits using \OIII\ as the narrow-line template for all sources that do not appear in Figure~\ref{spec_fitting}.

%
\begin{figure*}[h]
	\centering
	\figurenum{A1}
	\includegraphics[height=17.5cm]{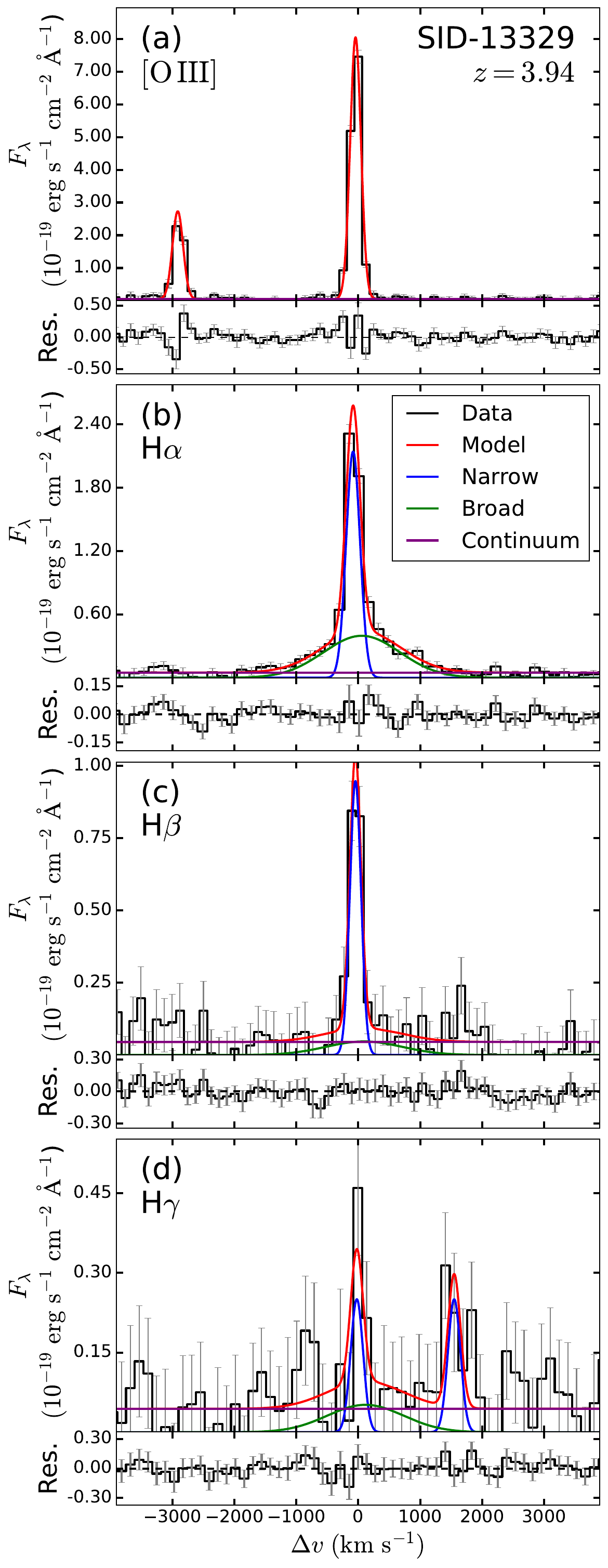}
	\includegraphics[height=17.5cm]{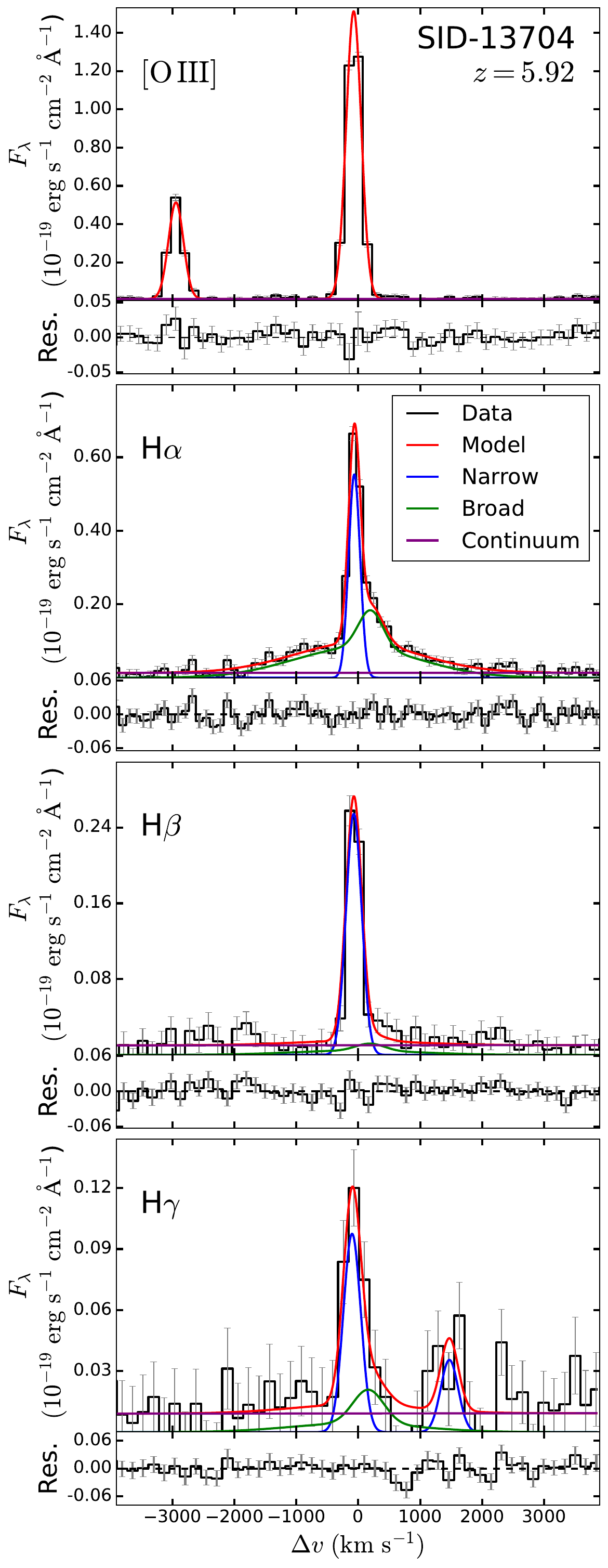}
	\caption{Spectral fits for (a) [\ion{O}{3}] $\lambda\lambda 4959, 5007$, (b) H$\alpha$, (c) H$\beta$, and (d) H$\gamma$ for SID-13329 and SID-13704. The top row in each panel shows the data and uncertainty in black line with gray errorbars, and the continuum, narrow-line component, broad-line component, and total model purple, blue, green, and red lines, respectively. Residuals are displayed in the bottom row.}
	\label{spec_fitting_13329}

\end{figure*}

\begin{figure}
	\centering
	\figurenum{A2}
	\includegraphics[height=17.5cm]{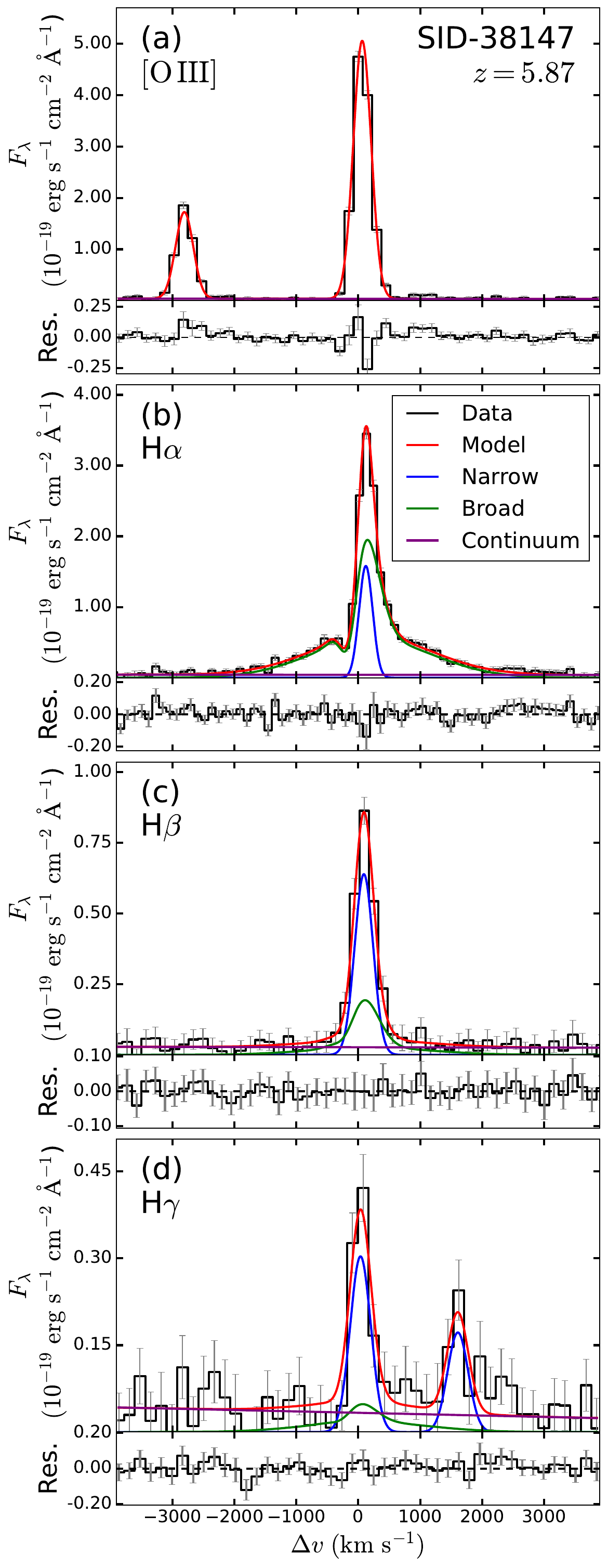}
	\includegraphics[height=17.5cm]{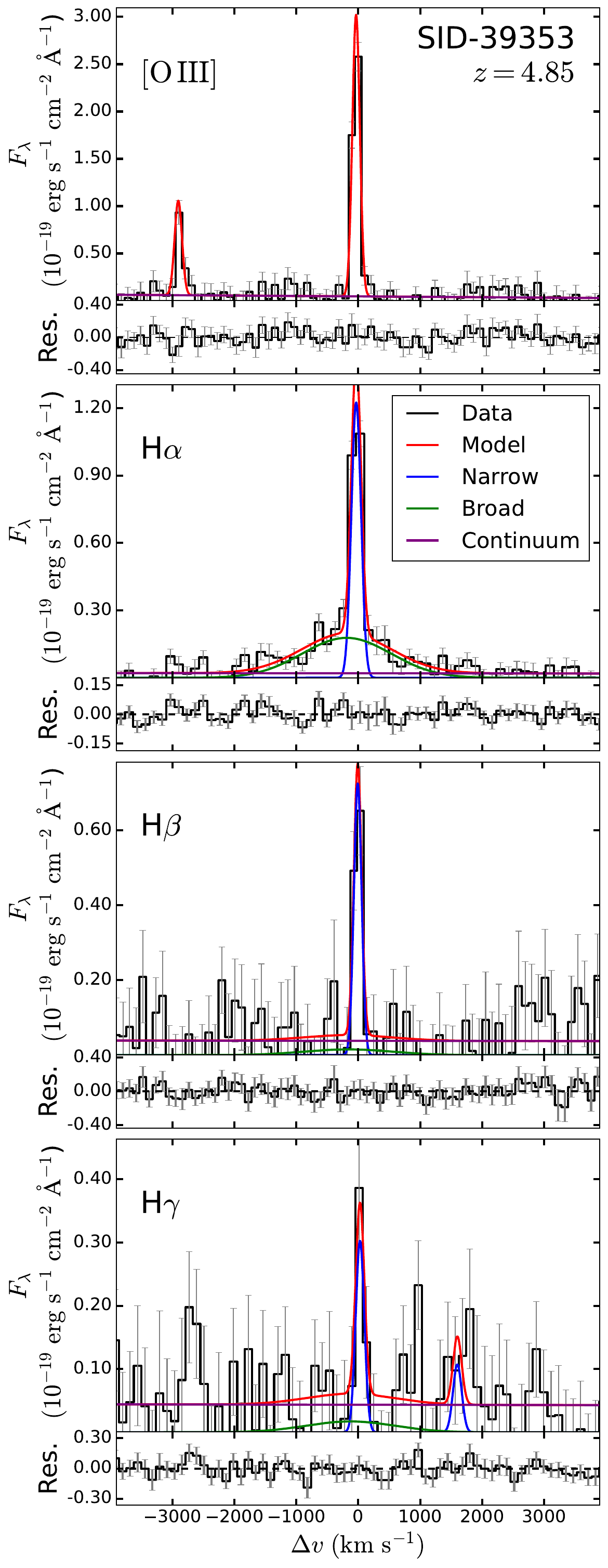}
	\caption{Same as Figure~\ref{spec_fitting_13329} but for SID-38147 and SID-39353. }

\end{figure}

\begin{figure}
	\centering
	\figurenum{A3}
	\includegraphics[height=17.5cm]{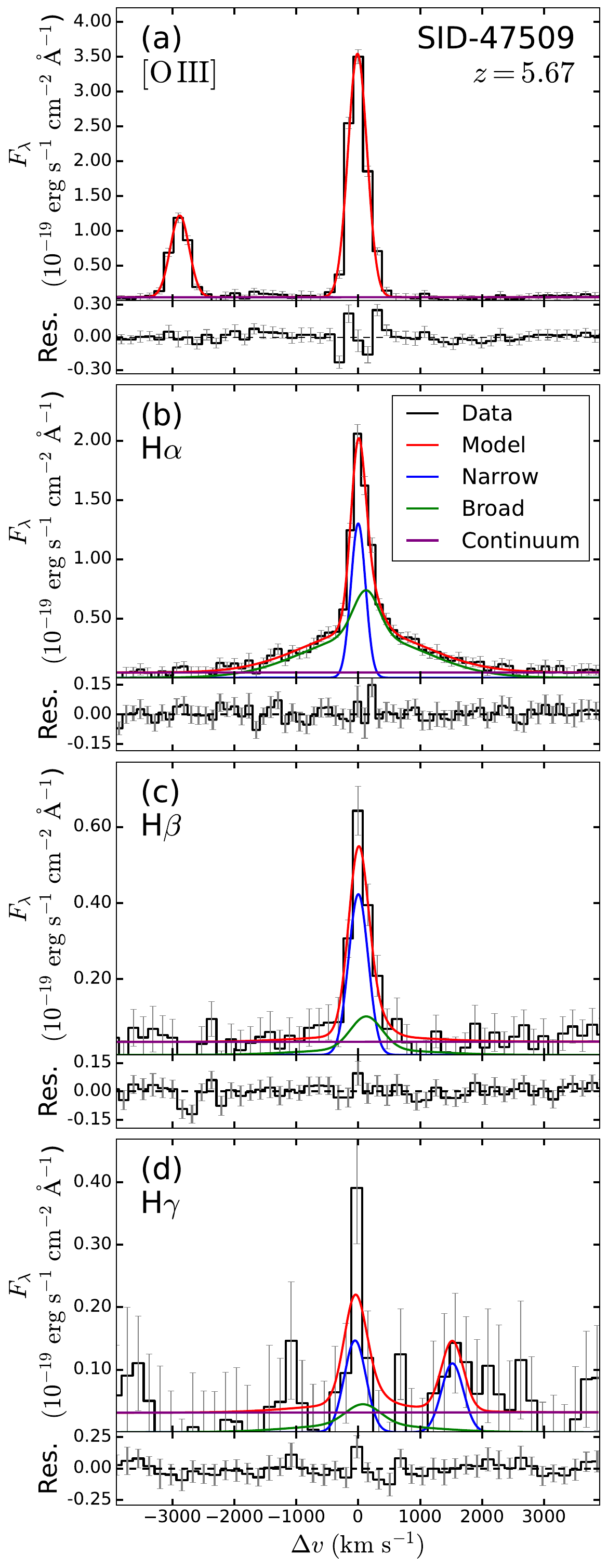}
	\includegraphics[height=17.5cm]{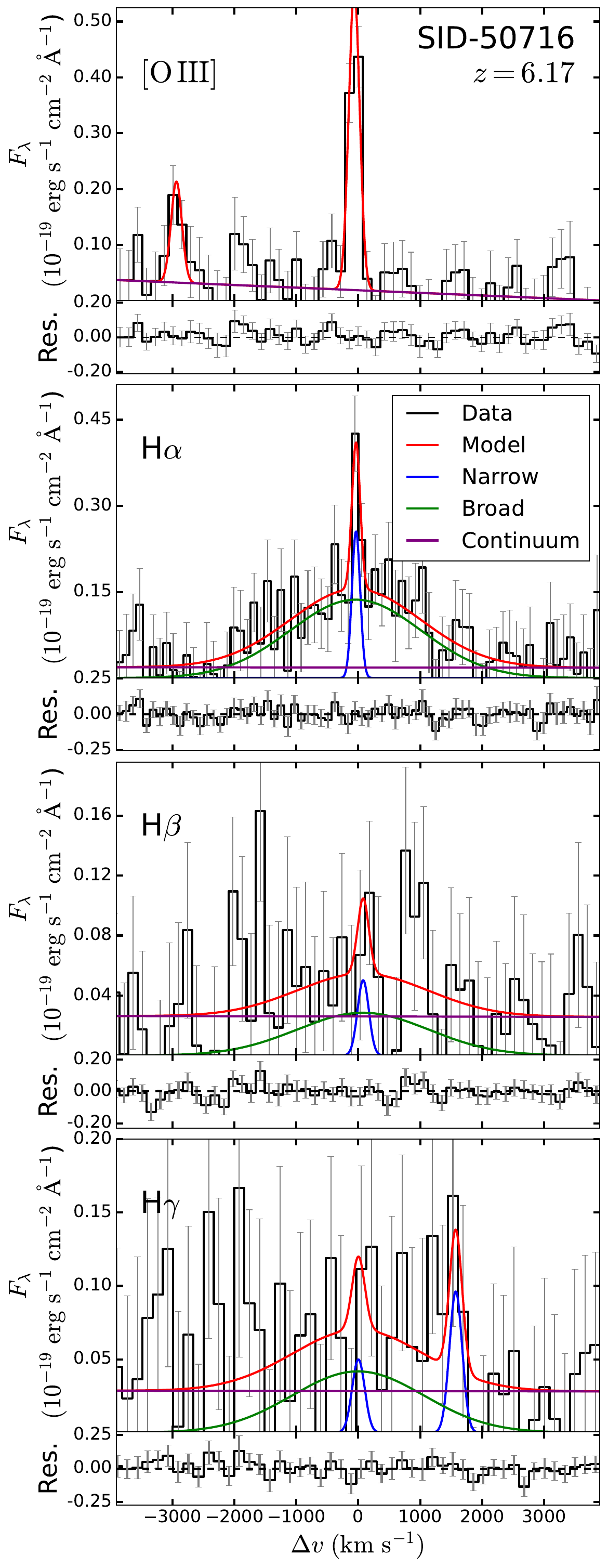}
	\caption{Same as Figure~\ref{spec_fitting_13329} but for SID-47509 and SID-50716. }

\end{figure}

\begin{figure}
	\centering
	\figurenum{A4}
	\includegraphics[height=17.5cm]{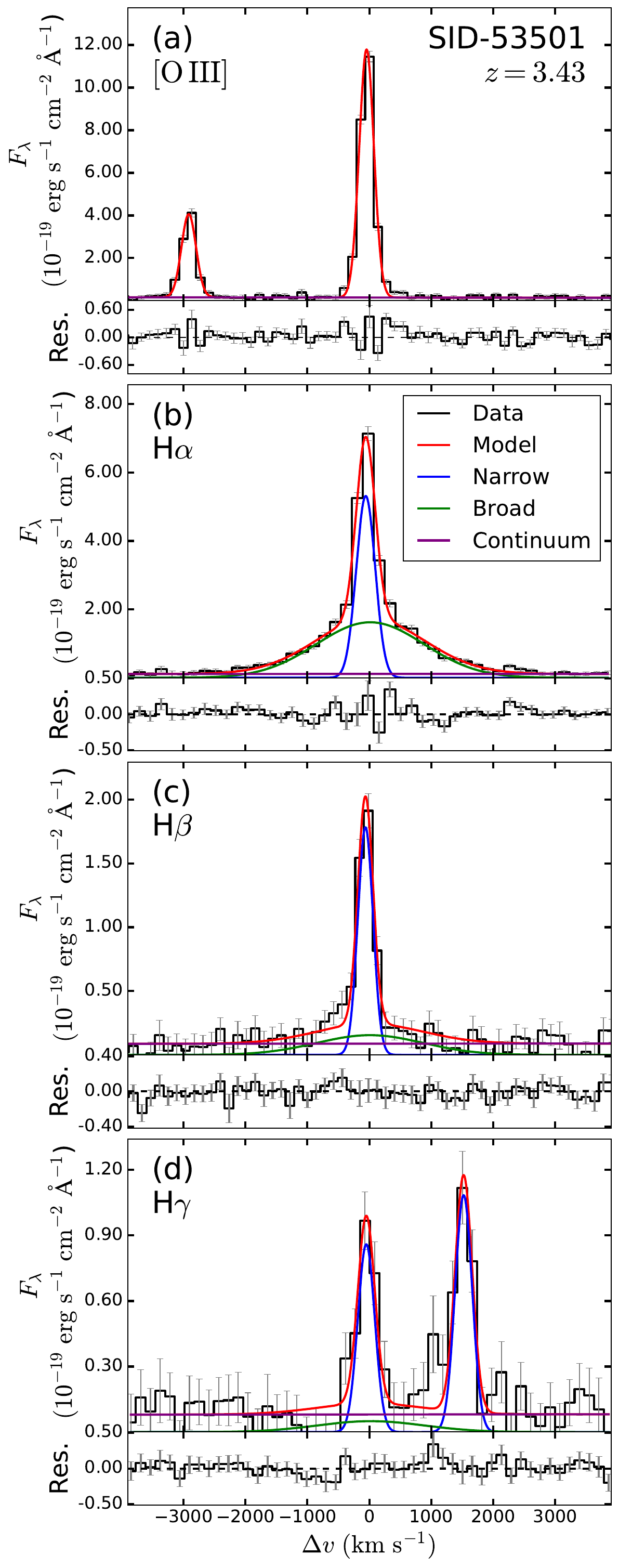}
	\includegraphics[height=17.5cm]{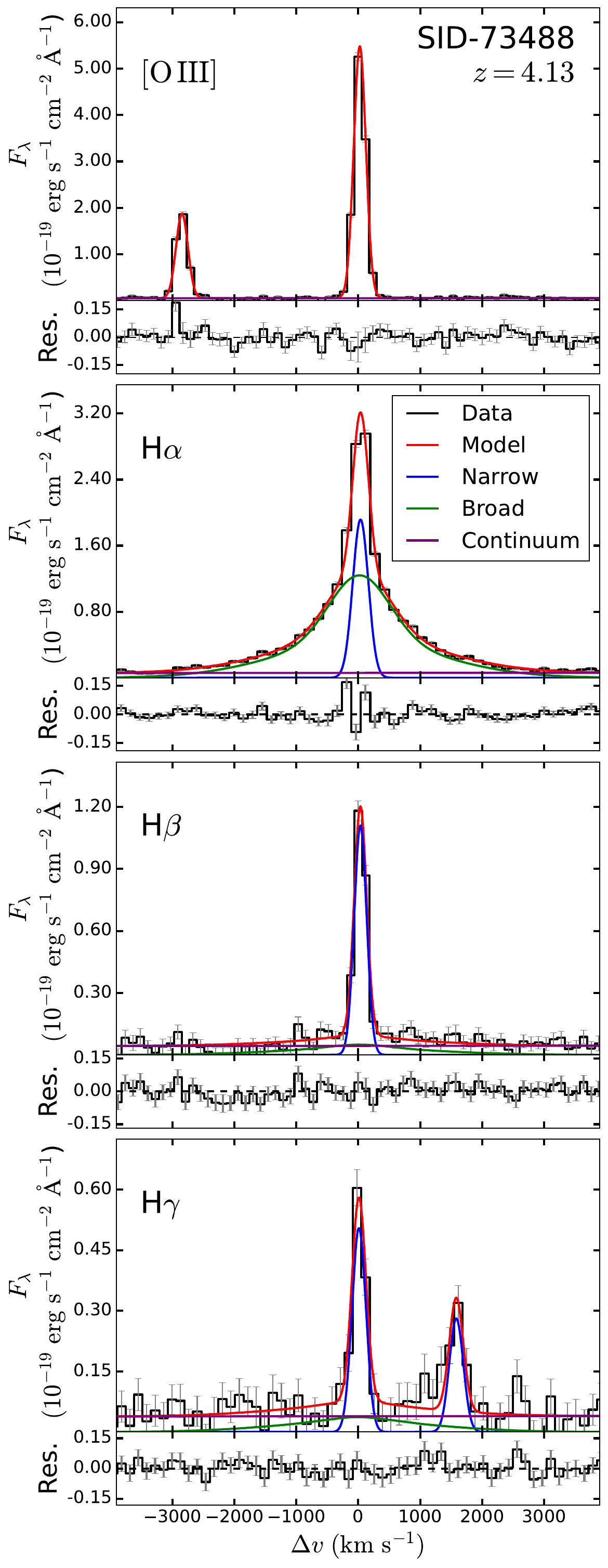}
	\caption{Same as Figure~\ref{spec_fitting_13329} but for SID-53501 and SID-73488. }

\end{figure}



\begin{figure*}[ht!]
    \centering
    \figurenum{A5}

    \begin{minipage}[t]{0.38\textwidth}
        \vspace{0pt}
        \includegraphics[height=13cm]{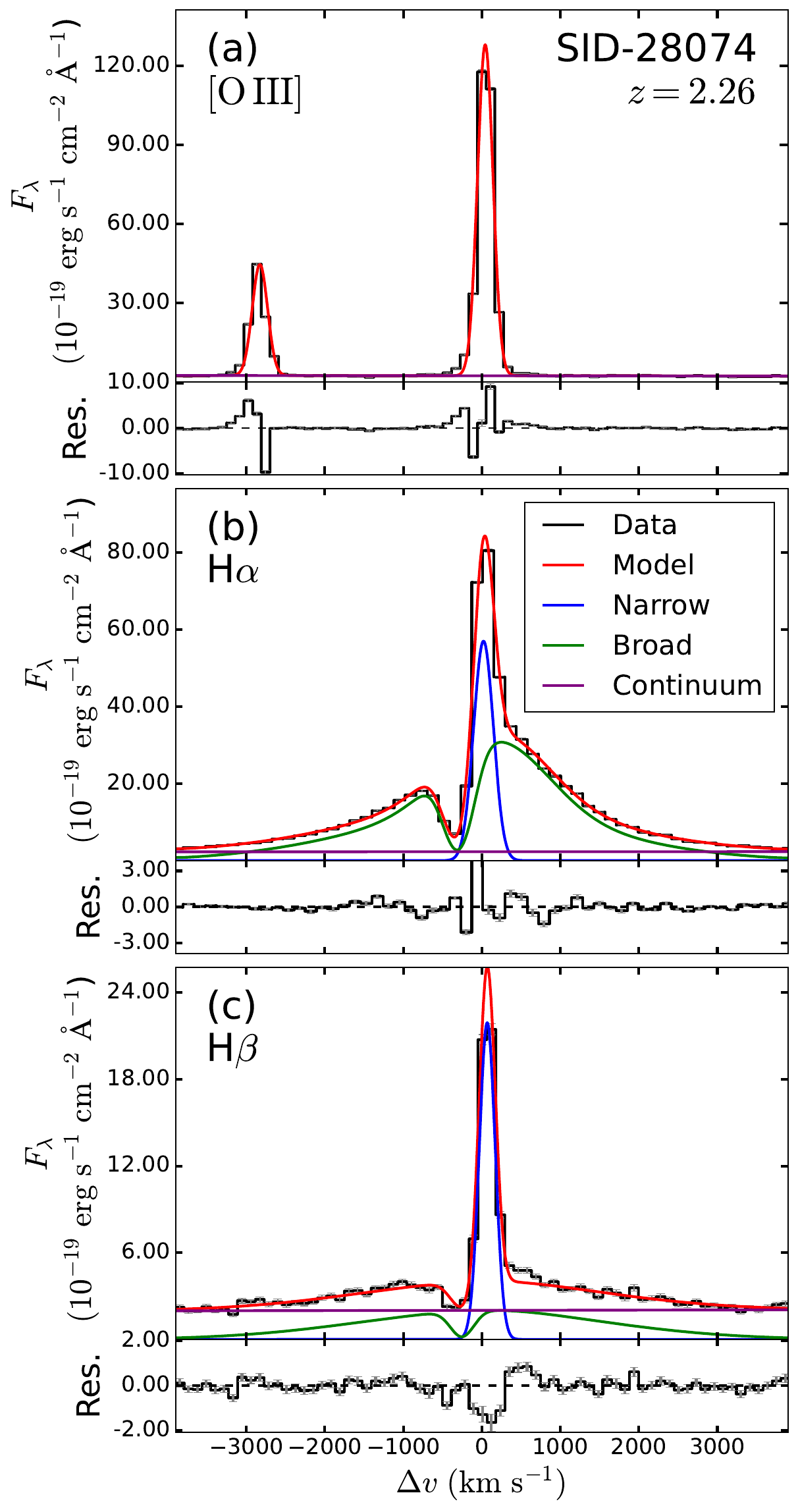}
        \hfill
    \end{minipage}
    \begin{minipage}[t]{0.38\textwidth}
        \vspace{0pt}
        \includegraphics[height=13cm]{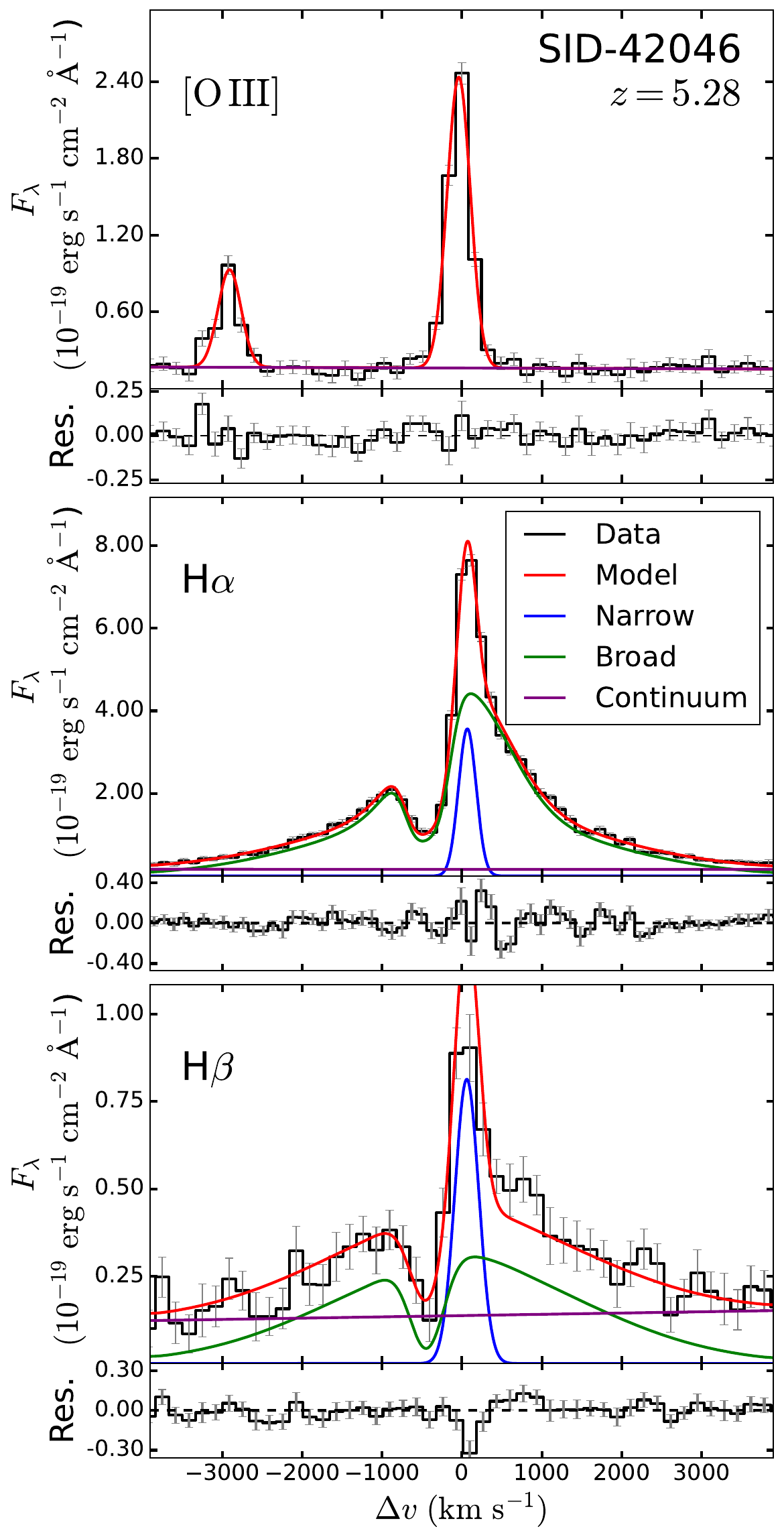}
        \hfill
    \end{minipage}

    \caption{Same as Figure~\ref{spec_fitting_13329} but for SID-28074, SID-42046 and SID-55604.}
\end{figure*}

\begin{figure}
	\centering
	\figurenum{A4}
	\includegraphics[height=13cm]{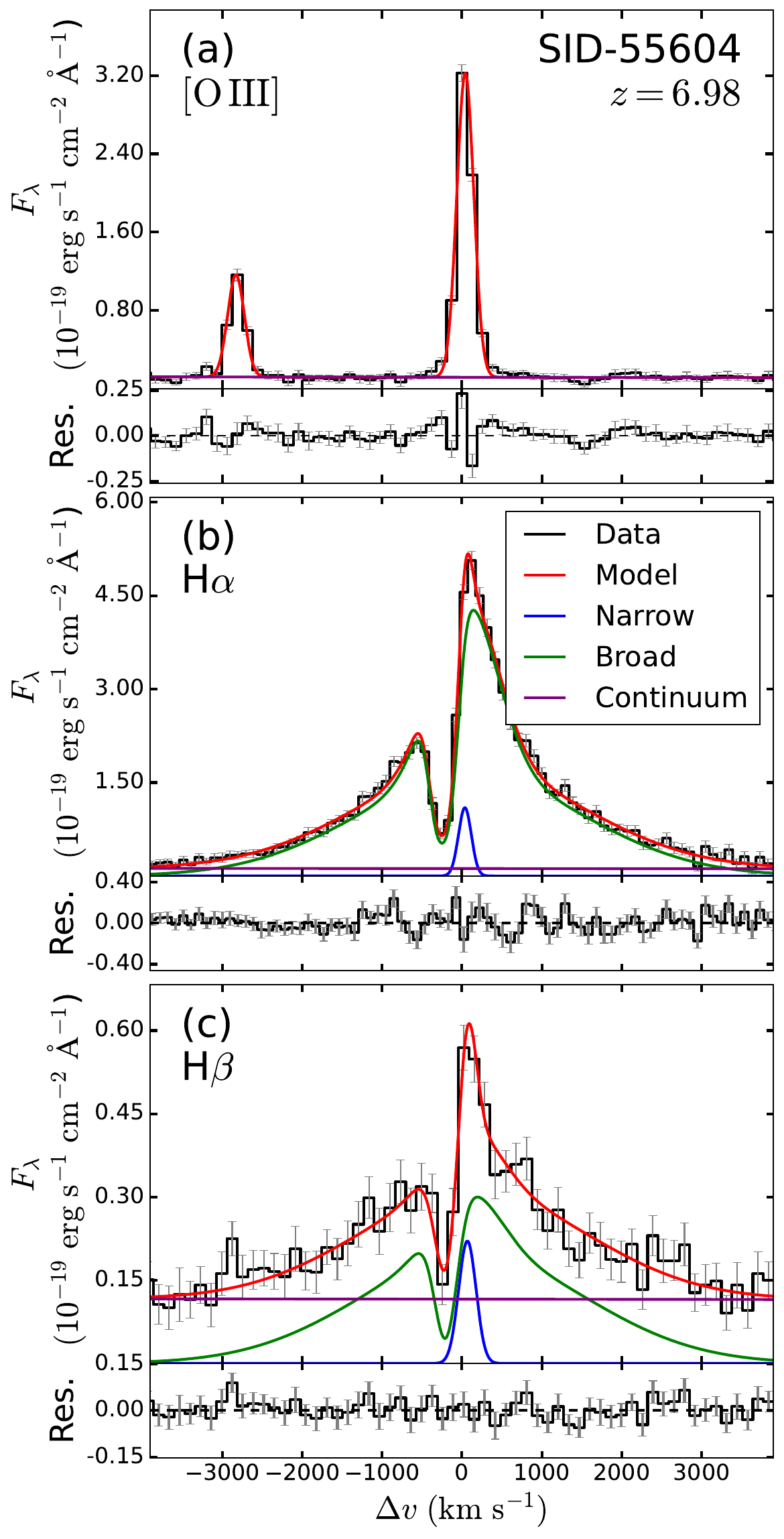}
	\caption{Same as Figure~\ref{spec_fitting_13329} but for SID-55604. }

\end{figure}

\clearpage

\section{Mock Absorption-line Fitting Results}

Fitting results of the mock spectra for sources with absorption lines detected. 

\begin{figure*}[ht!]
        \figurenum{B1}
	\centering
	\includegraphics[width=1\textwidth]{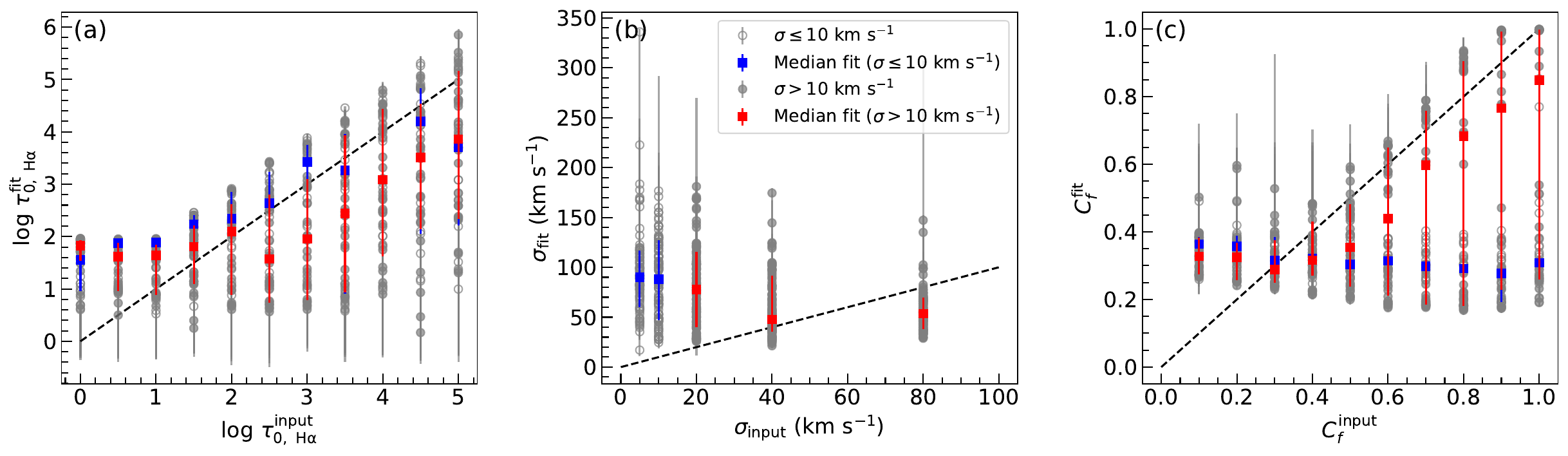}
	\caption{Same as Figure~\ref{abs_mock_42046} but for SID-28074.}
	\label{abs_mock_28074}
\end{figure*}

\begin{figure*}[ht!]
        \figurenum{B2}
	\centering
	\includegraphics[width=1\textwidth]{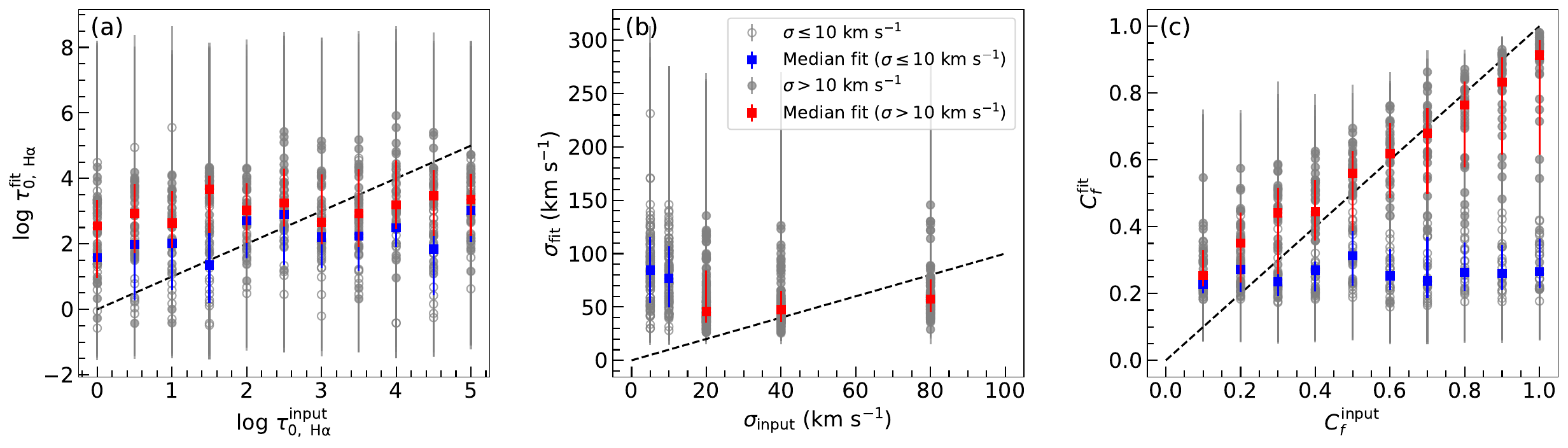}
	\caption{Same as Figure~\ref{abs_mock_42046} but for SID-38147.}
	\label{abs_mock_38147}
\end{figure*}

\begin{figure*}[ht!]
        \figurenum{B3}
	\centering
	\includegraphics[width=1\textwidth]{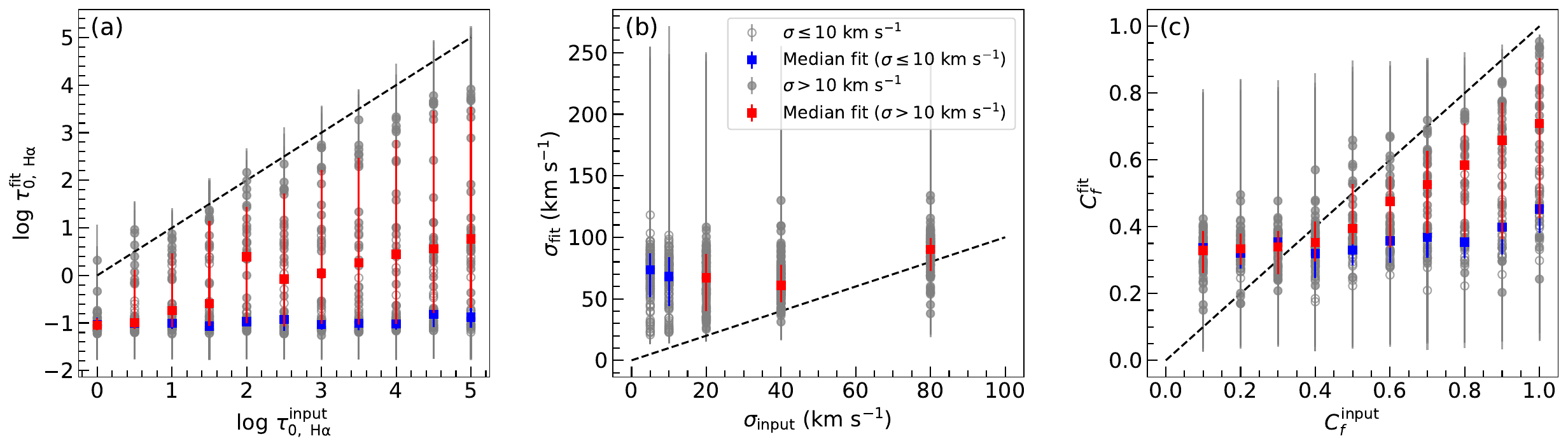}
	\caption{Same as Figure~\ref{abs_mock_42046} but for SID-49140.}
	\label{abs_mock_49140}
\end{figure*}

\begin{figure*}[ht!]
        \figurenum{B4}
	\centering
	\includegraphics[width=1\textwidth]{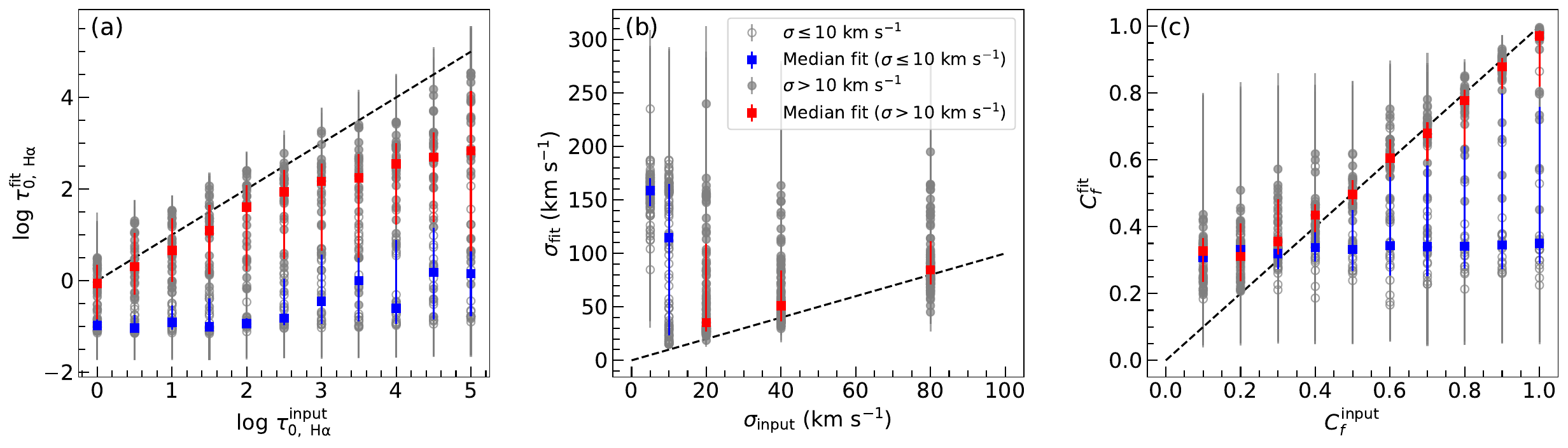}
	\caption{Same as Figure~\ref{abs_mock_42046} but for SID-55604.}
	\label{abs_mock_55604}
\end{figure*}

\begin{figure*}[h!]
        \figurenum{B5}
	\centering
	\includegraphics[width=1\textwidth]{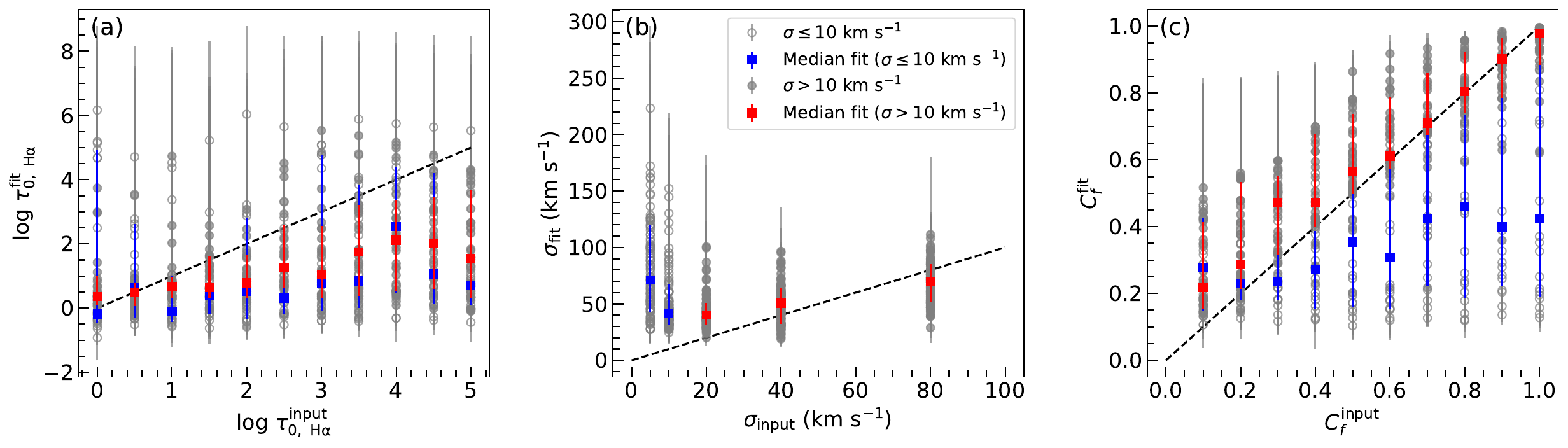}
	\caption{Same as Figure~\ref{abs_mock_42046} but for SID-68797.}
	\label{abs_mock_68797}
\end{figure*}

\clearpage
\newpage
\section{Exponential broadening and P-Cygni absorption model}\label{exp_pcygni}
 
We explore an alternative line-model combination for the five sources with blueshifted absorption features, three of which have unphysically high Balmer decrements, in order to assess the systematic uncertainties in the narrow-line measurements.
\citet{Naidu2025_bh_star} and \citet{Rusakov2025} proposed that the broad Balmer lines may not be broadened solely by virial motions but instead also by electron scattering in high-density gas. Under this scenario, an exponential profile is added to model the broad emission-line component, and the absorption trough is modeled using a P-Cygni profile, potentially originating from a hypothetical expanding gas shell.
	
Following the procedures in \citet{Nikopoulos2025} and \citet{Rusakov2025}, we fit the broad line with two components. A Gaussian component is used to represent the unscattered light that leaks out from the gaseous sphere, whose FWHM specifies the dynamical broadening in the vicinity of the BH. The scattered component is modeled by the convolution of the unscattered Gaussian component with an exponential function,
	
\begin{equation}
f=A\cdot {\rm exp}\left(-\frac{\left| v-v_0 \right|}{w} \right),
\end{equation}

\noindent
in which $A$ and $v_0$ are the amplitude and systemic velocity. The parameter $w$ stands for the e-folding scale of the exponential profile, which only depends on the electron temperature, as Thompson scatter is a frequency-independent, elastic process.  

We use the code from \citet{Sneppen2023}, written by Ulrich Noebauer\footnote{https://github.com/unoebauer/public-astro-tools}, to generate the P-Cygni profile, which is calculated based on the Elementary Supernova (ES) model described in \citet{Jeffery1990}. Assuming a homologous expansion sphere, the velocity between any two elements is proportional to the distance between the elements. Therefore, when a static sphere is considered, as is the case for the LRDs in our work, the spatial coordinate can be expressed in velocity at any given time. We fix the time to be $t=3000\ {\rm s}$, and the profile can be described by five parameters, $v_{\rm max}$, $v_{\rm phot}$, $v_{\rm ref}$, $v_{e}$ and $\tau_{\rm ref}$. Parameter $v_{\rm max}$ and $v_{\rm phot}$ stands for the velocity (position) of the outer and inner boundary of the sphere.  Parameter $v_{\rm ref}$ and $v_{e}$ describe the density profile for the gas between the inner and outer boundary, while $\tau_{\rm ref}$ normalizes the total optical depth of the sphere. Together, the optical depth distribution as a function of velocity (radius) is modeled as
	
\begin{equation}
    \tau(v)=\tau_{\rm ref}\cdot {\rm exp}\left( \frac{v_{\rm ref}-v}{v_{e}} \right).
\end{equation}

\noindent
Following \citet{Rusakov2025}, we fix $v_{e}$ to 500 ${\rm km~s^{-1}}$, varying $v_{\rm max}$, $v_{\rm phot}$, and $\tau_{\rm ref}$ to modify the shape of the absorption profile, and allowing $v_{\rm ref}$ to shift the centroid of the profile. The absorption model has four free parameters in total. The overall model can be described as

\begin{equation}
f_{m}=f_{n}+\left( f_{b}+f_{\rm exp}+f_{\rm cont} \right)\cdot R_{\rm cyg},
\end{equation}

\noindent
where $f_{m}$, $f_{n}$, $f_{b}$, $f_{\rm exp}$, and $f_{\rm cont}$ denote, respectively, the flux of the total model, the narrow-line component, the unscattered and scattered broad-line component, and the continuum. The P-Cygni profile is given by $R_{\rm cyg}$. We simultaneously fit \ha, \hb, and \hg, with the intrinsic FWHM of the unscattered component and the e-folding scale of the exponential profile tied across the three lines. For the P-Cygni absorption profile, we tie the parameter $v_{\rm max}$ and $v_{\rm phot}$, and we fix the ratio of the reference optical depth $\tau_{\rm ref}$ according to theoretical predictions. 

The fitting results are shown in Tables~\ref{results_table_exp} and \ref{results_table2_exp}. Within the context of the exponential broadening and P-Cygni absorption model, the narrow-line Balmer decrements in three sources (SID-28074, SID-42046, SID-55604) are still high, with measurement or upper limit larger than 4.7, while the narrow-line component in SID-38147 and SID-68797 is not detected at all. We compare the BIC of this alternative P-Cygni profile model with that of our fiducial Gaussian model for the five sources discussed here. The alternative model yields BIC values lower than those of the fiducial model by 30 to 140 for four sources (SID-28074, SID-42046, SID-55604, SID-68797), while the fiducial model gives BIC value lower by 30 for SID-38147. We note, however, that uncertainties in the LSF and the attenuation curve can plausibly affect the BIC at a level comparable to the differences between the models. A detailed comparison between these physical interpretations is beyond the scope of this work. We therefore conclude that the Balmer decrements of these sources, although uncertain, are moderately high. 

\begin{deluxetable*}{ccccccccccccccccccccc}[]
\label{results_table_exp}
\tabletypesize{\footnotesize}
\tablecaption{Emission-line Measurements Based on Exponential and P-Cygni Profile}
\tablewidth{0pt}

 \tablehead{ID & R.A. & Decl. & $z$ & \multicolumn{2}{c}{${\rm H\alpha/H\beta}$} & \multicolumn{2}{c}{${\rm H\gamma/H\alpha}$} & \multicolumn{2}{c}{${\rm log}\ ({L_{\rm inci}}\ {\rm erg\ s^{-1}})$} & \multicolumn{2}{c}{${\rm log}\ ({L_{\rm obs}}\ {\rm erg\ s^{-1}})$} \\
\cmidrule(lr){5-6} \cmidrule(lr){7-8} \cmidrule(lr){9-10} \cmidrule(lr){11-12}
& (deg) & (deg) & & narrow & broad & narrow & broad & narrow & broad & UV & optical \\                 
(1) & (2) & (3) & (4) & (5) & (6) & (7) & (8) & (9) & (10) & (11) & (12)}

\setlength{\tabcolsep}{5pt}
\startdata
28074 & $189.0646$ & $62.2738$ & 2.26 & $5.69_{-0.10}^{+0.10}$ & $12.18_{-0.59}^{+0.62}$ & \nodata & \nodata & $43.84_{-0.01}^{+0.01}$ & $44.89_{-0.20}^{+0.15}$ & $46.19_{-0.01}^{+0.01}$ & $44.89_{-0.08}^{+0.07}$ \\
38147 & $189.2707$ & $62.1484$ & 5.87 & \nodata & $12.14_{-3.25}^{+5.71}$ & \nodata & $0.03_{-0.01}^{+0.01}$ & \nodata & $43.39_{-0.28}^{+0.22}$ & $44.03_{-0.03}^{+0.03}$ & $43.35_{-0.06}^{+0.05}$ \\
42046 & $214.7954$ & $52.7888$ & 5.28 & $>4.74$ & $11.91_{-1.95}^{+1.92}$ & \nodata & \nodata & $43.37_{-0.02}^{+0.02}$ & $44.83_{-0.42}^{+0.28}$ & $45.33_{-0.05}^{+0.05}$ & $44.61_{-0.02}^{+0.02}$ \\
55604 & $214.9830$ & $52.9560$ & 6.98 & $7.21_{-2.59}^{+3.95}$ & $13.37_{-0.66}^{+0.72}$ & \nodata & \nodata & $43.29_{-0.06}^{+0.06}$ & $45.63_{-0.08}^{+0.08}$ & $45.94_{-0.10}^{+0.09}$ & $45.06_{-0.21}^{+0.16}$ \\
68797 & $189.2291$ & $62.1462$ & 5.04 & \nodata & $20.59_{-3.52}^{+3.98}$ & \nodata & $0.02_{-0.01}^{+0.01}$ & \nodata & $43.68_{-0.16}^{+0.15}$ & $43.86_{-0.02}^{+0.02}$ & $44.48_{-0.01}^{+0.01}$ \\
\enddata

\tablecomments{
Col. (1): Source identification from DJA.
Col. (2): Right ascension (J2000).
Col. (3): Declination (J2000).
Col. (4): Redshift.
Cols. (5)--(8): Line intensity ratios of the the narrow and broad components.
Cols. (9)--(10): Incident ionizing luminosity inferred from the narrow and broad components of \ha. 
Cols. (11)--(12): Observed UV and optical continuum luminosity.  }
\end{deluxetable*}

\begin{deluxetable*}{cccccccccc}[]
\label{results_table2_exp}
\tabletypesize{\footnotesize}
\tablecaption{Derived Physical Parameters}
\tablewidth{0pt}
\tablehead{ID & log $n_{\rm H}$ &  log $N_{\rm H}$ & $A_V$ & Break & log ($v_{\rm max}$) & log ($v_{\rm phot}$) & $\tau_{\rm ref}$ & log $M_{\rm BH}$ \\
& $({\rm cm^{-3}})$ & $({\rm cm^{-2}})$ & (mag) & Strength & (c) & (c) & & ($M_{\odot}$) \\
(1) & (2) & (3) & (4) & (5) & (6) & (7) & (8) & (9)}
\setlength{\tabcolsep}{5pt}
\startdata
28074 & $11.25_{-0.27}^{+0.20}$ & $23.34_{-0.10}^{+0.09}$ & $2.37_{-0.06}^{+0.06}$ & $2.26\pm0.25$ & $-2.50_{-0.01}^{+0.01}$ & $-2.90_{-0.02}^{+0.02}$ & $8.70_{-0.74}^{+0.82}$ & $7.76_{-0.09}^{+0.07}$ \\
38147 & "$9.30_{-0.46}^{+0.80}$ & $22.46_{-0.26}^{+0.26}$ & \nodata & $1.15\pm0.25$ & $-2.92_{-0.13}^{+0.08}$ & $>-2.84$ & $>1.91$ & $4.96_{-0.09}^{+0.08}$ \\
42046 & $10.64_{-0.75}^{+0.68}$ & $23.60_{-0.17}^{+0.10}$ & $>1.74$ & $3.55\pm0.46$ & $-2.40_{-0.01}^{+0.01}$ & $-2.62_{-0.03}^{+0.03}$ & $>27.33$ & $6.45_{-0.08}^{+0.07}$ \\
55604 & $11.77_{-0.10}^{+0.09}$ & $23.48_{-0.10}^{+0.07}$ & $3.19_{-1.54}^{+1.51}$ & $3.36\pm0.46$ & $-2.77_{-0.02}^{+0.02}$ & $-3.33_{-0.10}^{+0.09}$ & $>2.43$ & $7.70_{-0.09}^{+0.08}$ \\
68797 & "$8.62_{-0.10}^{+0.27}$ & $23.48_{-0.36}^{+0.36}$ & \nodata & $2.62\pm0.46$ & $-2.48_{-0.02}^{+0.01}$ & $-2.62_{-0.02}^{+0.02}$ & $>5.89$ & $5.56_{-0.09}^{+0.08}$ \\
\enddata
\tablecomments{
Col. (1): Source identification from DJA. 
Cols. (2)--(3) Volume density and column density measured from the Balmer break and Balmer decrement. 
Col. (4): Dust attenuation inferred from the narrow-line Balmer decrement.
Col. (5): Balmer break strength. 
Cols. (6)--(7): Line center optical depth of H$\alpha$ and the covering factor for the absorption component.
Cols. (8)--(9): Central velocity offset of the absorption line relative to the narrow emission line and the intrinsic velocity dispersion.
Col. (10): BH mass estimated using the single-epoch method of \citet{Greene_Ho_2005}.  }
\end{deluxetable*}

\end{CJK*}
\end{document}